\documentclass[11pt, a4paper, oneside, reqno]{amsart}
\usepackage{comment}
\usepackage[usenames, dvipsnames]{color}
\definecolor{darkblue}{rgb}{0.0, 0.0, 0.45}
\definecolor{lightblue}{RGB}{240,248,255}
\definecolor{lightblue2}{rgb}{0.68, 0.85, 0.9}
\definecolor{lightcyan}{rgb}{0.88, 1.0, 1.0}
\definecolor{palepink}{rgb}{0.98, 0.85, 0.87}

\usepackage[colorlinks	= true,
raiselinks	= true,
linkcolor	= darkblue, 
citecolor	= Mahogany,
urlcolor	= ForestGreen,
pdfauthor	= {Peyman Mohajerin Esfahani},
pdftitle	= {},
pdfkeywords	= {},
pdfsubject	= {},
plainpages	= false]{hyperref}

\usepackage{dsfont,amssymb,amsmath,enumitem} 
\usepackage{amsfonts,dsfont,mathtools, mathrsfs,amsthm} 
\usepackage[amssymb, thickqspace]{SIunits}
\usepackage{fancyhdr,mdframed,nicefrac}

\usepackage{epsfig}
\usepackage{graphicx}
\usepackage{float}
\usepackage{caption}
\usepackage{subcaption}

\usepackage{courier}

\usepackage{multirow}
\usepackage{bigstrut}

\allowdisplaybreaks
\date{\today}
\makeatletter
\def\@settitle{\begin{center}%
		\baselineskip14\p@\relax
		\normalfont\LARGE\scshape\bfseries
		\@title
	\end{center}%
}

\def\@setauthors{%
  \begingroup
  \def\thanks{\protect\thanks@warning}%
  \trivlist
  \centering\footnotesize \@topsep30\p@\relax
  \advance\@topsep by -\baselineskip
  \item\relax
  \author@andify\authors
  \def\\{\protect\linebreak}%
  \authors%
  \ifx\@empty\contribs
  \else
    ,\penalty-3 \space \@setcontribs
    \@closetoccontribs
  \fi
  \endtrivlist
  \endgroup
}

\makeatother
\makeatletter

\def\subsection{\@startsection{subsection}{2}%
	\z@{.5\linespacing\@plus.7\linespacing}{.5\linespacing}%
	{\normalfont\large\bfseries}}

\def\subsubsection{\@startsection{subsubsection}{3}%
	\z@{.5\linespacing\@plus.7\linespacing}{.5\linespacing}%
	{\normalfont\itshape}}

\usepackage{multirow}
\usepackage{bigstrut}
\usepackage{wrapfig}
\usepackage{caption}

\renewcommand{\geq}{\geqslant}

\renewcommand{\leq}{\leqslant}

\DeclareSymbolFont{symbolsC}{U}{pxsyc}{m}{n}

\usepackage{bm}
\usepackage{amsthm}

\usepackage{rotating}

\usepackage{algorithm}    
\usepackage{algpseudocode} 

\usepackage{tikz}
\usepackage[subnum]{cases}
\newcommand{\RNum}[1]{\uppercase\expandafter{\romannumeral #1\relax}}

\title[Multi-phase flow beyond equilibrium]{
Fokker-Planck-Poisson kinetics: \\Multi-phase flow beyond equilibrium
}
 \author{Mohsen Sadr, Marcel Pfeiffer, and M. Hossein Gorji}
 \thanks{Corresponding author: Mohsen Sadr}
 \thanks {Email: sadr@acom.rwth-aachen.de}
 \thanks{Mohsen Sadr: Applied and Computational Mathematics, RWTH Aachen University, Schinkestrasse 2, D-52062 Aachen, Germany. Marcel Pfeiffer: Institute of Space Systems, University of
Stuttgart, Pfaffenwaldring 29, D-70569 Stuttgart, Germany. M. Hossein Gorji: Laboratory of Multiscale Studies in Building Physics, Empa,  Swiss Federal Laboratories for Materials Science and Technology, D\"{u}bendorf, Switzerland.}
 
 \definecolor{ms1}{rgb}{0.0, 0.0, 0.0}
 \definecolor{ms2}{rgb}{0.0, 0.0, 0.0}
 \definecolor{ms3}{rgb}{0.0, 0.0, 0.0}
  \definecolor{ms4}{rgb}{0.0, 0.0, 0.0}
\date{June 16, 2021}

\begin{document}

\begin{abstract}
Multi-phase phenomena remain at the heart of many challenging fluid dynamics problems. Molecular fluxes at the interface determine the fate of neighboring phases, yet their closure far from the continuum needs to be modeled. Along the hierarchy of kinetic approaches, a multi-phase particle method is devised in this study. {\color{ms1} This approach is built closely upon the previous studies on the kinetic method-development for dense gasses  [Phys. Fluids, 29 (12), 2017] and long-range interactions [J. Comput. Phys, 378, 2019].
It is on this background that the current work on Fokker-Planck-Poisson modeling of multi-phase phenomena is initiated.} Molecular interactions are expressed via stochastic forces driven by the white noise, coupled to the long-range attractions. The former is local and pursues diffusive approximation of molecular collisions, whereas the latter takes a global feature owing to mean-field forces. The obtained Fokker-Planck-Poisson combination provides an efficient work-flow for physics-driven simulations suitable for multi-phase phenomena far from the equilibrium. Besides highlighting the computational efficiency of the method, various archetypical and complex problems ranging from inverted-temperature-gradient between droplets to spinodal decomposition are explored. Detailed discussions are provided on different characteristics of the droplets dispersed in low/high density background gases; including the departure of heat-fluxes from Fourier's law as well as droplets growth in spinodal phases.

\end{abstract}

\maketitle

\section{Introduction}
\noindent Conventional hydrodynamics can be extended in order to account for multi-phase phenomena. The Young-Laplace model for surface-tension besides the Maxwell equation for evaporation/condensation are typically employed. Along with Newton's viscosity law and Fourier's law of heat-fluxes, they form linear-response approximations for unclosed variables arising from conservation laws. It is well established that these closures become less reliable once strong departures from the equilibrium is observed, e.g. see \cite{worner2012numerical,schwarzkopf2011multiphase}. In practice, this continuum breakdown of multi-phase flows arises in near vacuum conditions and/or miniaturized devices. Plume-droplet interaction in space thrusters \cite{benson2004kinetic}, shale gas flows through ultra-tight porous media \cite{alharthy2013multiphase} and laser-droplet interaction in high-energy lithography \cite{hudgins2016neutral}, can be mentioned as few examples. \\ \ \\ 
The problem of modeling surface and flux terms can be circumvented by resetting conservation laws in the kinetic theory framework. The latter describes the dynamics by means of the molecular velocity distribution which accounts for the probability that a particle has a certain velocity at a given point in the space-time coordinate. Accordingly, various macroscopic field quantities can be recovered by conditional expectation of molecular quantities. The Boltzmann equation provides an accurate description of the velocity distribution in the dilute gas limit, subject to the molecular chaos assumption \cite{Bird,Chapman1953,Cercignani}. By  modifying the Boltzmann collision operator, the Enskog equation can be devised for dense gases  \cite{enskog1922kinetische,van1973modified,resibois1978h}. This is achieved by including the volume exclusion in the collision operator and adjusting the collision frequency. As the attractive forces of the molecular potential is represented by the mean-field theory, the Enskog-Vlasov (EV) equation is obtained \cite{vlasov1978many}. The EV equation can account for liquids, gases and multi-phase flows accurately  \cite{karkheck1981kinetic, grmela1971kinetic}. While the EV equation is adopted as the basis of this study, meso-scale models for multi-phase phenomena can be derived from other routes as well. For example, the Landau-Lifschitz fluctuating hydrodynamics offers a framework to introduce correlations among macroscropic fluxes and hence yields improvement over conventional hydrodynamics \cite{landau1959fluid}. Following this idea,  Smoothed-dissipative particle-hydrodynamics has been proposed for simulations of multi-phase flows beyond hydrodynamics \cite{espanol2003smoothed,liu2007dissipative}. 
\\ \ \\
Foremost, direct computation of the velocity distribution from kinetic equations should deal with the high-dimensionality of molecular degrees-of-freedom. Moreover, an accurate discretization of the collision operator has to honour collisional scales which yields stiffness issues near the equilibrium. Finally, long-range interactions described in the EV equation introduces yet another challenge for efficient computations. 
Spectral and discrete velocity methods have been pursued aiming at accurate solution of Boltzmann and Enskog equations (see e.g. \cite{broadwell_1964,gamba2009spectral,wu2013deterministic,wu2015fast,wu2016non}). The problem of high-dimensionality of kinetic systems can be relaxed by approximating the velocity distribution via finite number of moments. These so-called moment methods which are applicable for flows not too far from the equilibrium, have been devised using Grad's ansatz, Chapman-Enskog expansion, Maximum-Entropy solution and quadrature methods (see e.g. \cite{ kremer1988enskog,torrilhon2016modeling,struchtrup2019grad,wang2020kinetic,heylmun2019quadrature}). \\ \ \\ Alternatively, the curse of high-dimensionality can be avoided by evolving samples of the distribution according to the underlying stochastic process; at the cost of introducing statistical noise. Direct Simulation Monte-Carlo (DSMC) introduced by \cite{Bird1963} is most popular particle algorithm for rarefied gas flow simulations. By adopting same principles, 
 \cite{montanero1997simulation}, \cite{Alexander1995}, {\color{ms2} \cite{hadjiconstantinou2000surface},} \cite{Frezzotti1997} and \cite{frezzotti2019direct} extended DSMC algorithm for the dense flows. However as DSMC can successfully circumvent the high-dimensionality issue by resetting the kinetic system into the Monte-Carlo framework, it inherits stiffness of the collision operator near the equilibrium. This fundamental issue lies on the fact that the stochastic process resulting from the collision operator is not continuous in the velocity space. Thus time step sizes smaller than the mean-collision-time and spatial discretization smaller than the mean-free-path should be adopted. \\ \ \\
 The high-computational cost associated with dense collisions, motivated \cite{Jenny2010} to introduce a particle approach based on the Fokker-Planck (FP) approximation of the Boltzmann equation. The main idea behind the FP model is to describe the jump process resulting from the collision operator by an equivalent drift-diffusion process. Hence the resulting stochastic paths become continuous and thus the discretization constraints become more relaxed in comparison to DSMC type approaches. Further improvements of the FP model led to continuous stochastic description of dense gases on the basis of the Enskog equation  \cite{sadr2017continuous}. The FP model for dense gases (DFP) is constructed such that relaxation rates of various production terms of the Enskog collision operator,  as well as the collisional transfer of conserved quantities during the collision, are honoured.
\\ \ \\
At a different front, the computation of long-range interaction based on the mean-field limit can introduce further complexities due to long tail of the molecular potential. In contrast to kernel based estimation of long-range interactions \cite{frezzotti2005mean,piechor1994discrete,frezzotti2018mean}, a global solution algorithm can be devised as shown by \cite{sadr2019treatment}. First the attractive part of molecular potential can be approximated accurately via Green's function associated with an elliptic partial differential equation (PDE). The resulting screened-Poisson equation (SP) then can be adopted in order to approximate the Vlasov integral. Unlike density expansion methods which are local and require high regularity of the density field  \cite{he2002thermodynamic,korteweg1901forme}, the solution offered by the SP model retains  global features of the Vlasov integral and allows for strong variations of the density field. 
\\ \ \\
In this study, a solution algorithm is devised, which combines an accurate continuous description of the molecular collisions along with an efficient estimation of long-range interactions. The former is achieved by the DFP approximation of the Enskog equation, whereas the latter follows the SP model for the Vlasov integral. First in {
 \S~\ref{sec:rev_kinetic}, the kinetic theory as a mathematical model of multi-phase flows is described. In particular, the Enskog collision operator and the Vlasov integral are reviewed in \S~\ref{sec:enskog_operator} and \S~\ref{sec:vlasov_integral}, respectively. Then, the FP approximation of the Enskog equation along with the SP model of long-range forces are reviewed in \S~\ref{sec:DFP_model} and \S~\ref{sec:long-range}, respectively.}
 Next, an efficient particle Monte-Carlo solution algorithm for the Fokker-Planck-Poisson (DFP-SP) model suitable for multi-phase flows is assembled in \S~\ref{sec:algorithms_DFP_SP}.\\ \ \\
The resulting DFP-SP model then is validated in \S~\ref{sec:results}, by studying evaporation of droplets in {\color{ms4} equilibrium and }non-equilibrium settings. As the benchmark, numerical simulation based on the EV equation is obtained by combining the Enskog Simulation Monte-Carlo derived by \cite{montanero1997viscometric} for collisions, and Legendre-Gauss quadrature for the Vlasov integral. Next, the DFP-SP model is adopted to study inverted-temperature-gradients and spinodal decomposition problems in \S~\ref{sec:inverted} and \S~\ref{sec:spinodal_decomp}, respectively. For the former, the heat-transfer between two liquid slabs kept at a background gas is studied in a one-dimensional geometry. The results are compared with respect to benchmark EV solutions. By investigation of the single droplet evaporation, further insight is gained with respect to origin of inverted temperature gradient phenomenon. Finally gas-droplet decomposition in meta-stable setting known as the spinodal decomposition is studied in two and three dimensional settings. The simulation results are found to exhibit Lifschitz-Slyozov growth rate for the droplets. 
The validation study along with detailed simulation results provided for inverted temperature gradient and spinodal decomposition confirm efficiency and accuracy of the proposed DFP-SP model.
The paper concludes in \S~\ref{sec:conclusion}, where an outline of future improvements is discussed.
\section{Review of Enskog-Vlasov kinetic model}
\label{sec:rev_kinetic}
\noindent As number of molecules which constitute a typical fluid portion is immense, it is more natural to apply a statistical approach if we are interested in macroscopic behaviour of the system. The kinetic theory provides a description of the flow by means of the molecular distribution. Consider an ensemble of random particles with translational degrees-of-freedom. The state of the ensemble is fully determined by random variables $X(t)\in \mathbb{R}^3$ and $V(t)\in \mathbb{R}^3$ which account for position and velocity of a sample at time $t\in \mathbb{R}^+$, respectively. Equivalently, we can deduce the single-particle velocity distribution at position $x$ and velocity $v$ by
\begin{eqnarray}
\mathcal{F}(v,x,t)&=&\mathcal{M}\mathbb{E}\left[\delta(x-X(t))\delta (v-V(t))\right],
\end{eqnarray}
where $\mathbb{E}[.]$ is the expectation, $\delta (.)$ denotes the Dirac delta and $\mathcal{M}$ accounts for the total mass of the system (see e.g. \cite{Gorji2014a}). In absence of boundary effects, the distritbution $\mathcal{F}$ evolves due to the inter-molecular interactions. For our system of particles with molecular mass $m$, consider the Sutherland molecular potential
\begin{eqnarray}
\phi(r)=
\left\{\begin{array}{lr}
 +\infty & r<\sigma\\ 
 \phi_0 \Big (\dfrac{\sigma}{r} \Big )^6& r \geq \sigma
\end{array}\right. 
\label{eq:sutherland}
\end{eqnarray}
where $r$ is the distance between any two particles, $\sigma$ indicates the effective diameter and $\phi_0$ is a scaling factor. In principle for a given inter-molecular potential, the Liouville equation offers evolution of the multi-particle distribution as shown by \cite{cercignani1988boltzmann}. However if further simplifications are adopted i.e. the mean-field limit for long-range interactions along with the molecular chaos assumption, a closed form equation for $\mathcal{F}$ can be deduced \cite{grmela1971kinetic,karkheck1981kinetic,cercignani1988boltzmann}. Contributions of short-range interactions to the evolution of $\mathcal{F}$ can be captured by considering an elastic collision operator, whereas the $N$-body problem arising from the long-tail interaction can be approximated by the mean-field limit \cite{vlasov1978many}. Hence, the EV equation 
\begin{flalign}
\frac{\partial \mathcal{F}}{\partial t}
+ \frac{\partial (\mathcal{F} v_i)}{\partial x_i}
- \frac{1}{m} H_i \frac{\partial \mathcal{F}}{\partial v_i} 
= S^{\textrm{Ensk}}(\mathcal{F})
\label{eq:full_kinetic_model}
\end{flalign}
can be derived where the Enskog collision operator $S^{\textrm{Ensk}}(\mathcal{F})$ accounts for the contribution of molecular collisions. Furthermore, the conservative  force field $H_i :={\partial \Phi}/{\partial x_i}$ accounts for long-range attraction and follows the gradient of the mean-field potential $\Phi$. 
Here and henceforth the Einstein summation convention is used to economize our notation.
\\ \ \\
{\color{ms1}It is important to emphasize that the underlying assumptions behind the conventional EV equation makes it constrained to monatomic single species flows at moderate densities. However validation studies in comparison with Molecular Dynamics (MD) simulations in various non-equilibrium scenarios, motivates its further use and development for a broad range of multi-phase flow processes \cite{frezzotti1997molecular,kon2014method,frezzotti2008comparison,frezzotti2018mean}.}
\ \\ \\
In follow-up subsections, first we review the Enskog collision operator and next a short description on the mean field potential $\Phi$ is provided.
\subsection{Enskog operator}
\label{sec:enskog_operator}
The short-range interactions have two main influences on evolution of the velocity distribution: relaxation towards equilibrium and transport of conserved quantities. Note that the latter is only the case for dense flows.  The Enskog collision operator derived by \cite{enskog1922kinetische} and then modified by \cite{van1973modified,resibois1978h}, is an extension of the Boltzmann collision operator by including the volume of particles.
The standard Enskog collision operator for hard-sphere collisions reads
\begin{flalign}
S^{\text{Ensk}}(\mathcal{F})
= \frac{1}{m}& \int_{\mathbb{R}^{3}} \int_{0}^{2\pi} \int_0^\sigma   
 \Big [ Y ( {x}+\frac{1}{2} \sigma \hat{ {k}}) \mathcal{F}( {v}^*,  {x})\mathcal{F}( {v}^*_1,\ {x}+\sigma \hat{ {k}})
\nonumber \\
 &-Y ( {x}-\frac{1}{2} \sigma \hat{ {k}})\mathcal{F}( {v},  {x})\mathcal{F}( {v}_1,  {x}-\sigma \hat{ {k}}) \Big ] 
g  \mathcal{H}(  g_i \hat{ {k}}_i)  \hat{b} \ 
d\hat{b} d\hat{\epsilon}  d{v}_1~,
\label{eq:enskog_op}
\end{flalign}
where $\hat{k}$ indicates unit vector connecting the center of a particle to its colliding pair with subscript $(.)_1$, $\mathcal{H}(.)$  the Heaviside step function, $g$ is the relative velocity of colliding pair, and $\hat{b}$ and $\hat{\epsilon}$ are the impact parameter and scattering angle, respectively. Note that in case of hard-sphere potential, the impact parameter relates to deflection angle $\chi$ via $\hat{b}=\sigma \cos(\chi/2)$ where $\chi\in[0,\pi)$. For further details of the Enskog and Boltzmann collision operators see e.g., \cite{Chapman1953}, \cite{Hirschfelder1963} and \cite{Bird}. 
\\ \ \\ 
In comparison to the dilute limit, the Enskog collision operator admits a diameter distance between the colliding pair as well as an increase in collision rate by $Y$. The standard Enskog equation estimates the pair correlation factor $Y$ to be a function of density at the contact of colliding spheres, which can be chosen such that the correct equilibrium pressure is guaranteed. By means of the Carnahan-Starling equation of state, we get
\begin{flalign}
Y &= \frac{1}{2}\frac{2-\eta}{(1-\eta)^3}  \\
\text{with} \hspace{0.5cm} \eta &:= nb/4,
\label{eq:pair_corr}
\end{flalign}
where $b:=2 \pi \sigma^3/3$ is the second virial coefficient \cite{carnahan1969equation}. Furthermore, the number density is denoted by $n$ which gives the density $\rho=mn$.
\\ \ \\
\noindent In addition to the curse of dimensionality associated with $\mathcal{F}$, the increase of collision rate along with the non-locality of collision operator lead to further computational complications. Although direct Monte-Carlo particle methods can cope with the high-dimensionality at the cost of introducing statistical noise, they inherit the issues with non-locality of collision step along with resolving collisional scales. 
\subsection{Vlasov integral}
\label{sec:vlasov_integral}
By applying the mean-field limit, the attractive force of Eq.~\eqref{eq:full_kinetic_model} is represented as the derivative of the Vlasov integral
\begin{flalign}
\Phi(x,t) = \int_{r>\sigma} \phi(r) n(y,t) dy,
\label{eq:vlasov_integral}
\end{flalign}
where $r:=|y-x|$ and $| \ . \ |$ denotes Euclidean distance. Although the Vlasov integral can be computed using quadrature rules as discussed in Appendix~\ref{sec:num_long_range},  long-tails and mesh-refinement near $r=\sigma$ make the numerical integration expensive in practice. 
\section{Fokker-Planck-Poisson Model}
\noindent Here a detailed description of the Fokker-Planck-Poisson model covering both short- and long-range interactions is discussed.  First in \S~\ref{sec:DFP_model}, the FP approximation of collisions in dense gases is reviewed. Next, we explain the elliptic model that captures the long-range forces globally, by means of the SP equation in \S~\ref{sec:long-range}.
\subsection{Short-range interactions}
\label{sec:DFP_model}
\noindent The computational cost associated with resolving the Enskog collision operator   i.e. right hand side (rhs) of Eq.~\eqref{eq:full_kinetic_model}, can be decoupled from collisional scales once the underlying jump process is described by an equivalent continuous process. Furthermore, if the stochastic paths generated by the latter process depends only on local moments of the distribution, efficient cell-based parallelization can be exploited. 
\\ \ \\
Let us consider an  It\^{o} drift-diffusion process as a general prototype of such continuous process. Then, the task becomes to find a drift-diffusion closure such that the resulting single-particle velocity distribution evolves consistent with the underlying jump process, up to a desired degree of moments. From physical point of view such drift-diffusion process expresses all the forces acted on a sample particle into a portion which is correlated to the particle's state and a remainder being purely random. The former is governed by the drift whereas the latter is controlled by the diffusion coefficient.
\\ \ \\
Consider position and velocity of particles given by  random variables $(X(t),V(t))$. Suppose that evolution of particles can be modeled through  stochastic differential equations (SDEs) with drift $A$ and diffusion $D$ in velocity space, as well as a spatial drift $\hat{A}$ in the physical space
\begin{numcases}{}
  dV_i=  A_idt+D d  W_{t,i}\ \ \textrm{and}  \label{eq:sde_dfp_V} \\ 
  dX_i= V_idt + {\hat{A}_i} dt,
  \label{eq:sde_dfp_X}
\end{numcases}
where $i\in\{1,2,3\}$.
Here $dW_{t,i}:=W_i(t+dt)-W_i(t)$ indicates the increment of  the Wiener process for the $i$th component of particle velocity which follows a Gaussian distribution with zero mean and variance of $dt$, i.e. $dW_{t,i}\sim \mathcal{N}(0,dt)$. Using It\^{o}'s lemma, evolution of the corresponding $\mathcal{F}$ follow a FP equation of the form
\begin{flalign}
\dfrac{\partial \mathcal{F}}{\partial t}
+ \dfrac{\partial (\mathcal{F} v_i)}{\partial x_i}
= S^{\text{DFP}}(\mathcal{F}),
\label{eq:FP_eq}
\end{flalign}
where
\begin{flalign}
S^{\text{DFP}} (\mathcal{F})
= \underbrace{-\frac{ \partial (\mathcal{F} A_i )}{\partial v_i}
+\frac{1}{2}\frac{\partial^2 ( D^2 \mathcal{F})  }{\partial v_i \partial v_i}}_{\text{relaxation rates}} 
   \underbrace{-\frac{ \partial (\mathcal{F} \hat{A}_i )}{\partial x_i}}_{\text{col. transfer}}~.
\label{eq:FP_dense}
\end{flalign}
For relationship between the FP equation and SDEs see e.g. \cite{oksendal2013stochastic}.
The drift $A$ and diffusion $D$ allow us to fix the rate at which non-conservative velocity moments including kinetic stress-tensor
\begin{flalign}
\pi_{ij}^\mathrm{kin}  =\int_{\mathbb{R}^3} v_{\langle i}^\prime  v_{j \rangle}^\prime \mathcal{F} dv
\end{flalign}
and kinetic heat-flux vector 
\begin{flalign}
q_{i}^\mathrm{kin}  = \frac{1}{2} \int_{\mathbb{R}^3} v_i^\prime  v_j^\prime v_j^\prime \mathcal{F} dv,
\end{flalign}
relax due to collisions. Note that subscript $\langle . \rangle$ denotes the deviatoric part of the tensor and $v^\prime:=v-U$ is the fluctuating velocity with $ U=(1/\rho)\int   v \mathcal{F} d  v$ being the bulk velocity. \\ \ \\
The {\color{ms1}spatial} drift $\hat{A}$ accounts for the collisional transfer resulting from dense effects. From physical point of view, the introduced drift in the position-update corrects the particle position due to the volume exclusion. In other words, in absence of such correction the described SDEs only provide velocity and position of particles with vanishing diameter. 
\\ \ \\
By considering the spatially homogeneous setting, we fix $A$ and $D$ such that the resulting moment system remain consistent with the Enskog equation up to the heat-fluxes. Accordingly, the following constraints
\begin{flalign}
&\int_{\mathbb{R}^{3}} v^\prime_{\langle i} v^\prime_{ j \rangle} S^{\text{DFP}}(\mathcal{F}) d   {v}   =  \int_{\mathbb{R}^{3}} v^\prime_{\langle i} v^\prime_{ j \rangle} S^{\text{Ensk}}(\mathcal{F}) d   {v} \hspace{0.5cm} \text{and} \label{eq:relaxations_stress}\\
 &\int_{\mathbb{R}^{3}} v^\prime_{ i} v^\prime_{ j}  v^\prime_{ j}  S^{\text{DFP}} (\mathcal{F}) d   {v}  =  \int_{\mathbb{R}^{3}} v^\prime_{ i} v^\prime_{ j}  v^\prime_{ j}  S^{\text{Ensk}} (\mathcal{F}) d   {v}
\label{eq:relaxations_heat}
\end{flalign}
are devised.
In order to obtain a tractable expression for the rhs of Eq.~\eqref{eq:relaxations_stress}-\eqref{eq:relaxations_heat}, here we approximate moments of the Enskog collision operator by applying Maxwell molecules. Therefore we get
\begin{flalign}
&\int_{\mathbb{R}^3}  \left( A_i v^\prime_j + A_j v^\prime_i - \frac{2}{3} A_k v^\prime_k \delta_{ij}  + \frac{2}{3} \delta_{ij} D^2 \right) \mathcal{F} dv
 =  - Y \frac{p^{\mathrm{kin}}}{\mu^{\text{kin}} } \pi_{ij}^{\textrm{kin}} \ \ \ \textrm{and} \label{eq:relaxations_stress_enskog2}\\
 &\int_{\mathbb{R}^3}  \left( A_i v^\prime_j v^\prime_j + 2 A_j v^\prime_j v^\prime_i \right) \mathcal{F} dv  = - Y \frac{2}{3} \frac{p^{\mathrm{kin}}}{\mu^{\text{kin}} } q_i^{\textrm{kin}}~.
\label{eq:relaxations_heat_enskog2}
\end{flalign}
Note that here kinetic ideal gas viscosity coefficient is denoted by $\mu^{\text{kin}}$, ideal gas equilibrium  pressure is $p^{\text{kin}}=nk_b T$ and superscript $(.)^\mathrm{kin}$ indicates kinetic contribution.
\\ \ \\
Next, the extra transport resulting from spatial dependencies of the Enskog operator is accounted by the spatial drift. The degrees of freedom in extra streaming induced by $\hat{A}$ are set such that the resulting FP model remain consistent with the Enskog collision operator up to the heat-fluxes. Let $\psi=[1,v_i, v_jv_j/2]^T$, by the Taylor expansion of $\mathcal{F}(x\pm \sigma \hat{k})$ and $Y(x\pm \sigma \hat{k}/2)$ around $x$, the conservative moments of Enskog collision operator can be formulated as fluxes
\begin{flalign}
\int_{\mathbb{R}^3}& \psi_l S^{\text{Ensk}}(\mathcal{F}) d {v} 
= 
- \dfrac{\partial \Psi^\phi_{lk}}{\partial x_k}, \textrm{\ \ where}
\\
\Psi_{lk}^\phi &=
\frac{Y \sigma}{2m} \int  \int \int  \int  
 (\psi^*_l-\psi_l) 
  \mathcal{F} \mathcal{F}_1 
  \hat{k}_k g \hat{b} d\hat{b} d\hat{\epsilon}  d {v}_1 d {v}
  \nonumber \\
  &+
  \frac{Y \sigma^2}{4m} \int  \int   \int   
 (\psi^*_l-\psi_l)
\hat{k}_k \hat{k}_j  \mathcal{F} \mathcal{F}_1  \frac{\partial}{\partial x_j} \ln (\frac{\mathcal{F}}{\mathcal{F}_1})
g \hat{b} d\hat{b} d\hat{\epsilon}  d {v}_1 d {v}\nonumber \\
&+ \cdot \cdot \cdot ~.
\label{eq:psi_phi}
\end{flalign}
as shown by \cite{Chapman1953}.
In particular, the corrections accounted by the spatial drift $\hat{A}$ become
\begin{flalign}
\frac{\partial }{\partial x_k}
\int_{\mathbb{R}^{3}}  \hat{A}_k \psi_l \mathcal{F}  d   v \approx 
\dfrac{\partial \Psi^\phi_{lk}}{\partial x_k} 
\textrm{\ \ \ for }l=1,...,\dim(  \psi).
\end{flalign}
Assuming Maxwellian for $\mathcal{F}$ in $\partial_{x_j} \ln({\mathcal{F}}/{\mathcal{F}_1})$ inside the second integral of Eq.~\eqref{eq:psi_phi} and ignoring higher order terms, a closure for collisional transfer can be derived
\begin{flalign}
&\int_{\mathbb{R}^3} \hat{A}_i \mathcal{F} dv = 0,
\label{eq:mass_closure}
\\
\label{eq:pressure_closure}
&\int_{\mathbb{R}^3} \hat{A}_i v_j \mathcal{F} dv =  nbY(p^\mathrm{kin}\delta_{ij}+2/5\pi^\mathrm{kin}_{ij}) -w\left(\frac{\partial U_k}{\partial x_k}\delta_{ij}+\frac{5}{6}\frac{\partial U_{\langle i}}{ \partial x_{j \rangle}}\right)\\
\label{eq:heatflux_closure}
\text{and} \ \ 
 &\frac12 \int_{\mathbb{R}^3} \hat{A}_i v_j v_j \mathcal{F} dv =    \frac{3}{5} nbY   q_i^{\mathrm{kin}} - c_v w \frac{\partial T}{\partial x_i}.
\end{flalign}
Here $c_v=3 k_b /(2m)$ is the heat capacity at constant  volume,
\begin{eqnarray}
w &=& (nb)^2 Y \sqrt{mk_bT}/(\pi^{3/2} \sigma^2)
\label{eq:w}
\end{eqnarray}
is the second viscosity coefficient, $T= \int_{\mathbb{R}^3} v^\prime_j v^\prime_j \mathcal{F} dv/(3nk_b)$ is the temperature and $k_b$ denotes the Boltzmann constant.
\\ \ \\
A simple set of polynomial ansatz to fulfill Eqs.~\eqref{eq:relaxations_stress_enskog2}-\eqref{eq:relaxations_heat_enskog2} in combination with Eqs.~\eqref{eq:mass_closure}-\eqref{eq:heatflux_closure} is adopted.
Following \cite{sadr2017continuous}, the cubic FP model for dense collisions 
\begin{flalign}
A_i&=c_{ij}v^\prime_j+\gamma_i\left(v^\prime_jv^\prime_j-\frac{3k_bT}{m}\right)+\Lambda \left(v^\prime_iv^\prime_j v^\prime_j-\frac{2q^\mathrm{kin}_i}{\rho}\right),
\label{eq:dfp_vel_drift}\\
D &=\sqrt{\frac{k_bT}{\tau m}}  \\
\textrm{and}  \ \ \ \hat{A}_i &= \hat{c}_{ij} v^\prime_j + \hat{\gamma}_i \left(v^\prime_j v^\prime_j - \frac{3k_bT}{m}\right) 
+ \hat{\Lambda} \Big ( v^\prime_i v^\prime_j v^\prime_j -  \frac{2q^\mathrm{kin}}{\rho} \Big )
\label{eq:dfp_spa_drift}
\end{flalign}
is devised. 
The time scale $\tau$ is set using kinetic scales $\tau=2\mu^{\text{kin}}/(p^{\text{kin}}Y)$.
The coefficients of the tensors $c$ and $\hat{c}$ along with the vectors $\gamma$ and $\hat{\gamma}$ need to be computed based on velocity moments. The cubic term is responsible to make sure that the SDEs remain stable  (see \cite{Risken1989,Gorji2011}). The coefficients of the cubic terms are set to
\begin{flalign}
\Lambda&=-\frac{|\det (u_{ij})|}{ (u^{(2)})^4 \tau }\ \ \textrm{and}\\
 \hat{\Lambda} &= - \epsilon \frac{nbY}{k_b T/m }~,
\end{flalign}
where $\epsilon=10^{-3}$, $\det(.)$   indicates the determinant and
\begin{eqnarray}  
u_{i_1...i_n}^{(k)}&:=&\frac{1}{\rho}\int_{\mathbb{R}^{3}}| {v^\prime}|^kv^\prime_{i_1}v^\prime_{i_2}...v^\prime_{i_n}\mathcal{F}d   v~; \ \ \ \ i_k\in\{1,2,3\}
\end{eqnarray}
By substituting the ansatz for $A$ and $D$ in equations for the relaxation rates Eqs.~\eqref{eq:relaxations_stress_enskog2}-\eqref{eq:relaxations_heat_enskog2}, one can show that
\begin{flalign}
c_{ik}u^{(0)}_{kj}&+c_{jk}u^{(0)}_{ki}+\gamma_iu^{(2)}_{j}+\gamma_ju^{(2)}_{i}=-2\Lambda u^{(2)}_{ij}  \label{eq:vel_coeff1}\ \ \ \ \ \textrm{and} \\ 
c_{ij}u^{(2)}_j&+2c_{jk}u^{(0)}_{ijk}+\gamma_i(u^{(4)}-(u^{(2)})^2) 
+2\gamma_j(u^{(2)}_{ij}-u^{(2)}u_{ij}) \nonumber \\
&=-\Lambda (3u^{(4)}_i-u^{(2)}_iu^{(2)}-2u^{(2)}_ju^{(0)}_{ij})+\frac{5}{6}\frac{Yp^\mathrm{kin}}{\mu^{\text{kin}}}q_i^\mathrm{kin} \label{eq:vel_coeff2}.
\end{flalign}
Equivalently for $\hat{A}$ we get 
\begin{flalign}
\hat{c}_{jk}\pi_{ik}^\mathrm{kin}&+\hat{c}_{ji}p^\mathrm{kin}+2\hat{\gamma}_jq_i^\mathrm{kin}=-\rho \hat{\Lambda} u^{(2)}_{ij} \nonumber \\
&+nbY(p^\mathrm{kin}\delta_{ij}+2/5\pi^\mathrm{kin}_{ij})-w\left(\frac{\partial U_k}{\partial x_k}\delta_{ij}+\frac{5}{6}\frac{\partial U_{\langle i}}{ \partial x_{j \rangle}}\right) \ \ \textrm{and}
\label{eq:const_stress_total_DFP_SP}
\\
 \hat{c}_{ij} q_j^\mathrm{kin} 
           &+\frac{1}{2}\rho \hat{\gamma}_i    ( u^{(4)} 
           			   -   (u^{(2)}) ^2)  
           			    =\frac{3}{5}nbYq_i^\mathrm{kin} - w c_v\frac{\partial T}{\partial x_i} 
           			    - \frac{1}{2}\rho \hat{\Lambda}  (u^{(4)}_i
           			    - u^{(2)}_i  u^{(2)}   ),
\label{eq:const_heat_total_DFP_SP}
\end{flalign}
where $i,j,k\in\{1,2,3\}$. The unknowns $c,\ \gamma,\ \hat{c}$ and $\hat{\gamma}$, can be found by solving two linear systems 
\begin{flalign}
  A_v \begin{pmatrix}
 c_{1,1}\\ 
 \vdots\\ 
 c_{3,3} \\ 
 \gamma_{1}\\ 
 \vdots\\
 \gamma_{3}
\end{pmatrix}
=  b_v~
\ \ \mathrm{and} \ \ 
  A_x \begin{pmatrix}
 \hat{c}_{11}\\ 
 \vdots\\ 
 \hat{c}_{3,3} \\ 
 \hat{\gamma}_{1}\\ 
 \vdots\\
 \hat{\gamma}_{3}
\end{pmatrix}
=  b_x,
\label{eq:sys_vel_pos}
\end{flalign}
where the matrices $  A_v$ and $  A_x$ together with rhs matrices $  b_v$ and $  b_x$ are read from Eqs.~\eqref{eq:vel_coeff1}-\eqref{eq:vel_coeff2} and Eqs.~\eqref{eq:const_stress_total_DFP_SP}-\eqref{eq:const_heat_total_DFP_SP}, respectively. 
{\ \\ \ \\ \color{ms3} As shown in \cite{sadr2017continuous}, the  DFP model besides reproducing near-equilibrium results, accurately captures non-equilibrium effects within the transition regime for moderate densities. Specifically, we expect a reasonable accuracy for DFP model in the range of density where $nb\leq 1$.}

\subsection{Long-range interactions}
\label{sec:long-range}
\noindent The contribution of the long-range molecular potential in the evolution of molecular velocity distribution function can be accurately accounted by the Vlasov mean-field limit. Although the direct numerical evaluation of long-ranged forces through the Vlasov integral is grid-independent however it can lead to high-computational cost due to relatively large support of the kernel. In practice, global solutions over a spatial discretization turn out to be less  expensive \cite{sadr2019treatment}. Consider an approximation $\tilde{\phi}(r)$ to the attractive part of molecular potential $\phi(r)$ where
\begin{flalign}
\tilde{\phi}( r) &= a G( r) \hspace{0.5cm} \text{and}\\
G( r) &= \dfrac{e^{-\lambda r}}{4 \pi r}~.
\end{flalign}
An estimate of fitting parameters $a$  and $\lambda$ can be obtained by solving an optimization problem of the type
 \begin{flalign*}
(a,\lambda) &= \underset{r \in (\sigma, \infty)}{\operatorname{argmin}}( || \partial_r \phi(r)-\partial_r \tilde{\phi}(r)   ||_2^2)~
\end{flalign*}
for a given molecular potential. 
Here, $||.||_2$ indicates the $L^2$ norm. This ansatz is motivated by the fact that $G(r)$ is the Green function of the SP equation
\begin{eqnarray}
\label{eq:screen-unb}
\left(\Delta-\lambda^2\right)\tilde{\Phi}_{r>0}( x,t)&=&n( x,t); \ \ \ \ (\forall  x\in \mathbb{R}^3),
\end{eqnarray}
in three dimensional space. As shown above, the parameters $a$ and $\lambda$ give us flexibility to fit $\tilde{\phi}(r)$ to our desired potential. Hence, by substituting $\tilde{\phi}(r)$ in the Vlasov integral we obtain
\begin{flalign}
\label{eq:integ-sp}
\Phi( x, t)   &=  \int_{r>\sigma}\phi(  r) n( y, t) d  y \nonumber \\
&\approx  a\underbrace{\int_{r>0}G( r) n( y, t) d  y}_{\tilde{\Phi}_{r>0}}
   -   a\underbrace{\int_{r<\sigma}G( r) n( y, t) d y}_{\tilde{\Phi}_{r<\sigma}}\nonumber \\
   &= a\left( \tilde{\Phi}_{r>0} - \ \tilde{\Phi}_{r<\sigma} \right).
\end{flalign}
Further by assuming that the number density is smooth within $r\in(0,\sigma)$, an explicit expression for $\tilde{\Phi}_{r<\sigma}$ can be derived. In particular, using change of variables
\begin{flalign}
\tilde{\Phi}_{r<\sigma} (x,t) &= \int_{r<\sigma} G(|  y -   x|) n(  y, t) d^3   y\nonumber \\
&=   \int_{|  y|<\sigma} G(|  y| ) n(  y- x, t) d^3   y~,
\end{flalign}
and deploying Taylor's expansion for $n(y-x,t)$ around $(x,t)$, it can be shown that
\begin{flalign}
\tilde{\Phi}_{r<\sigma} (x,t)= & n(  x,t)  \underbrace{\int_{|  y|<\sigma} G(|  y| )d^3   y}_{\mathrm{analytical\  expr.}}
+ \frac{1}{2}\frac{\partial^2 n (  x,t)}{\partial x_j \partial x_j}   \underbrace{\int_{|  y|<\sigma} G(|  y|) y_j y_j d^3   y}_{\mathrm{analytical\  expr.}} +...\nonumber \\
\approx &
n(x,t) \left[e^{- \lambda \sigma} {\left(- \lambda \sigma + e^{\lambda \sigma} - 1\right) }/{\lambda^{2}}\right]\nonumber \\
&+ \frac{\partial^2 n(x,t)}{\partial x_j \partial x_j} \left[ e^{- \lambda \sigma} {\left(- \lambda^{3} \sigma^{3}/6 - \lambda^{2} \sigma^{2}/2 - \lambda \sigma + e^{\lambda \sigma} - 1\right) }/{\lambda^{4}}\right]~
\end{flalign}
by second order truncation. \\ \ \\
Therefore, instead of solving the Vlasov integral Eq.~\eqref{eq:vlasov_integral} directly, we find solution of the SP equation globally where the evaluated number density is fed as the source term. Fast Poisson solvers can be adjusted to the SP problem in the bounded domain $\Omega_x \subset \mathbb{R}^{3}$ with Dirichlet boundary conditions that provide the unique solution. The values of the potential on the boundaries of the domain $\partial \Omega_x$ are computed using the kernel $G(r)$ along with the boundary condition of $\mathcal{F}$. The detailed numerics of the SP model is presented \S~\ref{sec:discret_SP}. 
{\ \\ \ \\\color{ms3} As a simpler alternative to the SP model, one can obtain an explicit approximation of the Vlasov integral by expanding the number density around the point of interest and truncating the higher order terms. Although simple expressions can be obtained, this approach requires strong regularity assumptions behind the number density and also introduces further error as long-range contributions of the potential are computed locally, see \cite{sadr2019treatment}.}

\section{Particle method for Fokker-Planck-Poisson model}
\label{sec:algorithms_DFP_SP}
\ \\
Here we discuss a stochastic particle method for path simulations based on the devised FP approximation of the Enskog collision operator. Then, a discretization of the Screen-Poisson equation is provided. A solution algorithm combining both is outlined at the end.

\subsection{Stochastic particle method}
\label{sec:numerics_DFP}
\ \\
Since the SDEs \eqref{eq:sde_dfp_V}-\eqref{eq:sde_dfp_X} do not admit closed form time integration (due to non-linearity of the drift), a numerical scheme has to be adopted. 
As explained by \cite{Jenny2010,Gorji2011, Gorji2014}, simple Euler-Maruyama scheme does not guarantee energy conservation on average. Given the fact that the Langevin equation with linear drift allows for explicit time integration, a decomposition of the SDEs into a linear drift portion and a remainder is pursued. 
\\ \ \\
Following \cite{sadr2017continuous}, we consider a spatial discretization where each cell $c$ contains $N_{p,{(c)}}$ particles.  Then the evolution of each particle is described via explicit time-stepping scheme, where the solution at time step $(n+1)$ is computed based on the observables estimated at $(n)$th step. 
\\ \ \\
Realizations of the SDEs~\eqref{eq:sde_dfp_V}-\eqref{eq:sde_dfp_X},  through particles with weights $w^{(k)}$, velocities $  V^{(k)}$ and positions $  X^{(k)}$ with $k\in\{1,...,N_p\}$, provide an estimate
\begin{eqnarray}
\label{eq:dist-est}
\mathcal{F}(v,x,t)&\approx &\sum_{k=1}^{N_p}w^{(k)} \delta \left(  V^{(k)}(t)-  v\right)\delta\left(  X^{(k)}(t)-  x\right)
\end{eqnarray}
of the distribution which can be used to obtain various observables.
By generalizing particle discretization introduced by \cite{Gorji2014}, velocity and position of each particle get updated as
 \begin{numcases}{}
    V_i^{(*)}=  \langle V_i \rangle^{(n)}_{(c)}  + \alpha_{(c)}\tilde{V}^{(n)}_i
    \label{eq:vel_updates_DFP}
    \\
    X_i^{(n+1)}=   X_i^{(n)} + V_i^{(n+1)}\Delta t+\delta \tilde{X}^{(n)}_i
    \label{eq:pos_update_DFP}
  \end{numcases}
where
\begin{eqnarray}
\tilde{V}^{(n)}_i&=& 
 V_i^{\prime, (n)} e^{-\Delta tY^{(n)}/\tau^{(n)}} + c^{(n)}_{ij} V_j^{\prime, (n)} 
+ \gamma^{(n)}_i ( V_j^{\prime, (n)} V_j^{\prime, (n)} - 3k_bT^{(n)}/m ) 
\nonumber \\
&+&    \Lambda ( V^{\prime, (n)}_i V^{\prime, (n)}_j V^{\prime, (n)}_j -  2q^{(n)}_i/\rho^{(n)})
+ \sqrt{\frac{kT^{(n)}Y^{(n)}}{\tau^{(n)} m}\left({1-e^{\frac{-2\Delta tY^{(n)}}{ \tau^{(n)}}}}\right)} \xi_i
 \label{eq:FP_dense_velocity_update_DFP_SP} 
\end{eqnarray}
and
\begin{eqnarray}
\delta \tilde{X}^{(n)}_i=  \Big( \hat{c}^{(n)}_{ij} V^{\prime, (n)}_j &+& \hat{\gamma}^{(n)}_i (V^{\prime, (n)}_j V^{\prime, (n)}_j - {3k_bT^{(n)}}/{m}) 
\nonumber \\
&+& \hat{\Lambda}^{(n)} ( V^{\prime, (n)}_i V^{\prime, (n)}_j V^{\prime, (n)}_j -  2q^{(n)}_i/\rho^{(n)} 
\Big) \Delta t.
\label{eq:FP_dense_extra_streaming_part} 
\end{eqnarray}
Notice that $  V^{(*)}$ is modified accordingly once external or long-range forces are considered; otherwise we have $  V^{(n+1)} =   V^{(*)}$. Here  $\xi_i$ is a random number drawn from the standard normal distribution, i.e., $\xi_i \sim \mathcal{N}(0,1)$, and  $\alpha^{(c)}$ is the scaling factor which guarantees conservation of kinetic energy at each cell for a given time step
\begin{eqnarray}
\alpha_{(c)} = 
\frac{\langle V_{j}^{\prime, (n)} V_{j}^{\prime, (n)} \rangle_{(c)} }
{\langle \tilde{V}_j^{(n)} \tilde{V}_{j}^{(n)} \rangle_{(c)}}~.
\label{alpha_cell_DFP_SP}
\end{eqnarray}
Note that $\langle \ .\  \rangle^{(c)}$ indicates a moment estimator. Using Eq.~\eqref{eq:dist-est} we observe that 
\begin{flalign}
&\langle \hat{\phi} \rangle_{(c)} = \frac{1}{\hat{\rho}} \sum_{i\in \mathcal{S}_c} w^{(i)}{\hat{\phi}}^{(i)},
\label{eq:estimating_moments}
\end{flalign}
where ${\hat{\phi}}^{(i)}$ indicates $\phi$ evaluated by $i$th particle, $\mathcal{S}_c$ is set of particles in the cell $(c)$ and 
\begin{eqnarray}
\hat{\rho}&=&\frac{1}{\delta V_c}\sum_{i\in \mathcal{S}_c}w^{(i)}
\end{eqnarray}
is the estimated density based on the cell volume $\delta V_c$.
\\ \ \\
In comparison to conventional Langevin simulations, here the macroscopic coefficients $\{   c,   \gamma,  { \hat{c}},  { \hat{\gamma}}\}$ which close the drift term are computed from linear systems \eqref{eq:sys_vel_pos} for each computational cell and per each time step. Hence  correct relaxation rates in the velocity update besides accurate collisional transfer in the position update are obtained.
\subsection{Discretization of the screened-Poisson model}
\label{sec:discret_SP}
In order to translate the long-range interactions into a suitable elliptic problem, first we need to represent Eq.~\eqref{eq:screen-unb} in a bounded domain  $\Omega_x \subset \mathbb{R}^3$. Following uniqueness of the solution, the SP equation can be reset in a bounded domain $\Omega_x $ as
\begin{flalign}
\left\{\begin{matrix}
(\Delta -\lambda^2) \tilde{\Phi}_{r>0} &= -n  &\text{ \ in }\Omega_x   \\ 
\tilde{\Phi}_{r>0} &= \tilde{\Phi}_D  &\text{on }\partial \Omega_x,
\end{matrix}\right.
\label{eq:sp_bounded}
\end{flalign}
where the Dirichlet condition $\tilde{\Phi}_D$ is evaluated via 
\begin{flalign}
\tilde{\Phi}_D(x) = \int_{r>0} G(r) n(y) dy, \ \ \forall x\in\partial \Omega_x.
\label{eq:bc_sp}
\end{flalign}
The boundary term is computed numerically using quadrature estimation of the Vlasov integral and introducing a cut-off to contain the computational cost, i.e., $r\in (0, r_{\mathrm{cut}})$. Note that the singularity in $G(r)$ is avoided once the integral is formulated in the spherical coordinates. Here, we have set $r_{\mathrm{cut}} = 3 \sigma$ and used $N_{\text{quad}}$ Legendre-Gauss quadrature points for the numerical integration. Once the values of the unique global solution on the boundary $\partial \Omega_x$ is computed, this elliptic differential equation on the bounded domains can be solved for the interior of the domain using standard numerical methods. In \S\ref{sec:HDGS}, we present a short description for the adopted Hybridized Discontinuous Galerkin solver. Once our estimate of the attractive force at time step $(n)$ is obtained, the velocity update takes the form
\begin{eqnarray}
V^{(n+1)}&=&V^{(*)}+\hat{H}^{(n)}(X^{(n)})\Delta t,
\label{eq:up_vel_attraction}
\end{eqnarray}
where $\hat{H}^{(n)}(X^{(n)})$ is the approximated attractive force at the particle position $X^{(n)}$ (see Appendix \S \ref{sec:HDGS} for details).
\subsection{Solution algorithm}
\label{sec:description_algorithm}
\noindent It is evident from time-stepping scheme given by Eqs.~\eqref{eq:vel_updates_DFP}-\eqref{eq:pos_update_DFP} that the evolution of position and velocity of particles is decoupled during each time step. While more accurate schemes such as those discussed by \cite{Jenny2010,Gorji2011,jenny2019accurate} can be devised for a joint position-velocity update, here for simplicity the decoupled version is utilized. Note that as shown by \cite{pfeiffer2017adaptive,jun2019comparative}, this decoupling is accurate and justified as long as  the gradients of macroscopic fields are resolved. In comparison to DSMC \cite{Bird}, where the collisional scales have to be resolved, here we obtain a significant computational advantage by relaxing the discretization constraints. \\ \ \\
Algorithm \ref{alg:part_PF} provides a summary of the DFP-SP solution method for simulations of multi-phase flows. Prior to the simulation, the domain is discretized into a set of computational cells. Furthermore an appropriate time step size and particle weight are chosen. After initializing the particles according to their weights and initial density field, the particles evolve according to Alg.\ref{alg:part_PF}. Note that particles boundary conditions are implemented similar to DSMC, where depending on the distribution of incoming or reflected particles, open or wall boundaries are handled, respectively.     
\begin{algorithm}
\caption{DFP-SP solution algorithm}
\label{alg:part_PF}
\begin{algorithmic}
\While{$t<T$}
\State{Compute moments at each computational cell using Eq.~\eqref{eq:estimating_moments}.}
\State{Numerically compute the integral  Eq.~\eqref{eq:bc_sp} at the boundaries of the domain.}
\State{Solve linear systems in \eqref{eq:sys_vel_pos} for each cell.}
\State{Solve the global linear system Eq.~\eqref{eq:sp_bounded} for the whole domain.} 
\State{Compute $V^{(*)}$ of all particles based on stochastic process Eq.~\eqref{eq:vel_updates_DFP}.}
\State{Update the estimated velocity by including attractive forces Eq.~\eqref{eq:up_vel_attraction}.}
\State{Stream position of all particles, including dense effects with Eq.~\eqref{eq:pos_update_DFP}.}
\State{Apply the boundary conditions.}
\State{Increment $t$ by $\Delta t$.}
\EndWhile
\\
\Return
\end{algorithmic}
\end{algorithm}
\section{Validation results}
\label{sec:results}
\noindent For validation of the devised DFP-SP solution algorithm, we focus on evaporation of liquid Argon slab once in contact with the equilibrium vapor and once in contact with the vacuum. The coefficients of the Sutherland potential $\phi(r)$ are chosen according to $\sigma=3.405\times 10^{-10}\  \mathrm{m}$ and $\phi_0=4.17 \epsilon$ where $\epsilon = 119.8 k_b\  \mathrm{kg.m^2.s^{-2}} $ (provide citation). Fitting $\tilde{\phi}(r)$ to the attractive part of the Sutherland potential leads to the coefficients  $a=-1.64835851\times 10^{-28}\ \mathrm{kg.m^2.s^{-2}}$ and $\lambda=6.91304716\times 10^{9}\  \mathrm{m}^{-1} $.
\\ \ \\
We examine accuracy of the DFP-SP model in predicting capillary properties of droplets, such as the surface tension and the evaporation rate. The results are then compared with  benchmark Monte-Carlo simulations of the Enskog-Vlasov equation (ESMC-Vlasov). For implementation details of the benchmark ESMC-Vlasov approach see Appendix \S\ref{app:ESMC_Vlasov_alg}. 
\\ \ \\
The time step size satisfying a Courant-Friedrichs-Lewy (CFL) type condition
\begin{flalign}
\Delta t &\leq \frac{\min(\lambda^{(0)}, h)}{\max( \sqrt{k T^{(0)}/m})}\ , \label{eq:CFL_cond}\\
\textrm{where} \ \ \ \ \ \lambda^{(0)} &= \frac{1}{\sqrt{2}\pi\sigma^2n^{(0)} Y(n^{(0)})}
\end{flalign}
is chosen.
Here, $\lambda$ provides an estimate of the mean free path, $h$ indicates the mesh size, and superscript $(.)^{(0)}$ denotes the local initial value.
For further details see \cite{pfeiffer2017adaptive}. 
{\color{ms4}
\subsection{Isothermal co-existing states}
\label{sec:coex}
\noindent In this section, we investigate the accuracy of DFP-SP model in predicting the co-existing liquid-vapour states at constant temperatures against ESMC-Vlasov simulation results, besides the Maxwell area reconstruction solution \cite{li2013lattice}. We impose a thermostat every $N_\mathrm{therm}$ steps until the solution reaches stationary state, in order to keep the temperature of the system constant \cite{frezzotti2005mean}. The thermostat is then removed and afterwards the system reaches the equilibrium. The thermostat phase provides us with a decent initial distribution of particles in physical space for a given temperature. Furthermore, note that the solution after thermostat phase is already very close to the steady-state equilibrium solution.
\\ \ \\
In order to achieve simulation results of co-existing states, we consider a liquid slab $[0,1] \times [0,40\sigma] \times [0,1]\ \mathrm{m}^3$ with number density $n_l^{(0)}$ inside vapour with number density $n_v^{(0)}$. Here, the total volume $[0,1] \times [0,L] \times [0,1]\ \mathrm{m}^3$ with $L=100 \sigma$ is enclosed in a box with specular walls. Due to symmetry of the solution space, we only need to simulate the particle motion in the $x_2$ direction, yet velocity is treated in three dimensional space. The spatial dimension is decomposed to $100$ cells and $\Delta t$ was chosen to satisfy a CFL number of $0.01$. Initially, in each cell $N_{p/c}$ particles per cell with Gaussian distribution in velocities, i.e.,  $ V_i \sim \mathcal{N}(v_i|0,3k_bT^{(0)}/m)$, and uniform distribution in position are sampled. The statistical weight of particles $w$ was taken constant and chosen such that $N_{p/c}\geq 300$ in the vapour phase. The velocity of particles are  thermostated every $N_\mathrm{therm}=10$ steps for the first $1'000$ steps. Then, the simulation is continued without thermostat until stationary state is achieved ($\sim 2'000$ steps). Then, solution over next $3'000$ steps are averaged for the post-processing.
\begin{table}
  \begin{center}
\def~{\hphantom{0}}
  \begin{tabular}{cccccccccc}
      $T^{(0)}\ [\mathrm{K}]$ & $130$& $140$ & $150$ &$160$ &$165$ & $170$ & $175$ & $180$ & $185$
      \\[3pt]
       $n_v^{(0)}\sigma^3\ [-]$   & $0.0001$ &$0.0004$ & $0.003$ &   $0.009$ &    $0.03$ &    $0.06$ &   $0.08$ &    $0.12$ &    $0.16$\\
      $n_l^{(0)}\sigma^3\ [-]$   & $0.68$ & $0.64$ &   $0.58$ &   $0.54$ &    $0.52$ &   $0.48$ &    $0.44$ &    $0.4$ &    $0.35$
  \end{tabular}
  \caption{\color{ms4}Initial number density and temperature of liquid slab and vapour for the co-existing state simulations.}
  \label{tab:init_coex}
  \end{center}
\end{table}
\noindent Here, we consider several equilibrium states within $T \in [130,185]\ \mathrm{K}$. The initial number density of vapour and liquid for each case is provided in Table~\ref{tab:init_coex}. The density and temperature profiles for each case are shown in Figs.~\ref{fig2}-\ref{fig4}, respectively, where a reasonable consistency between  DFP-SP and ESMC-Vlasov solutions is observed. The obtained $(n,T)$ pair for liquid and vapour of each case is plotted against the Maxwell area reconstruction solution in Fig.~\ref{fig5}, where a reasonable agreement between the kinetic solution and Maxwell area reconstruction can be seen for its range of validity.
\\ \  \\
Note that the Maxwell solution leads to a minimization problem, i.e., find the pressure $p_0 = {\operatorname{argmin}} ( \int_{n_v}^{n_l} p(n,T)-p_0\  dn)$ for a given temperature. This pressure leads to two equal separated areas enclosed between $p(n,T)$ and the constant $p(n_l,T)=p(n_v,T)=p_0$. In order to compute the integral, one needs to find  $n_l$ and $n_v$ for any guessed pressure $p_0$ which leads to finding roots of a nonlinear equation $p(n,T)=p_0$ for $n$. In order to simplify computation of the Maxwell solution, we approximate the pair correlation function with a polynomial $Y(n)\approx 1 + 0.625n b + 0.2869 (n b)^2 + 0.115  (n  b)^3+...$, instead of Eq.~\eqref{eq:pair_corr}, see \cite{Chapman1953}. As shown in Fig.~\ref{fig1}, the polynomial approximation has a reasonable accuracy for the range of $n \sigma^3\leq 0.45$. Hence, we consider the equilibrium pressure 
\begin{flalign}
p_\mathrm{eq}^\mathrm{Max., Suth.} &= nk_bT(1+nb\tilde{Y}(n)) - \frac{4 \pi}{3} \phi_0 \sigma^2 n^2,\\
 p_\mathrm{eq}^\mathrm{Max., SP} &= nk_bT(1+nb\tilde{Y}(n))  + {a }  { e^{-\lambda \sigma} }{  }(\lambda^2 \sigma^2 + 3 \sigma \lambda +3 ) n^2/(6 \lambda^2)
 \ \ \mathrm{where}\\
\tilde{Y}(n) &= 1 + 0.625n b + 0.2869 (n b)^2 + 0.115  (n  b)^3,
\end{flalign}
for the Maxwell area reconstruction solution of co-existing states with validity range of $0<n \sigma^3\leq 0.45$.
\begin{figure}
    \centering
    \includegraphics{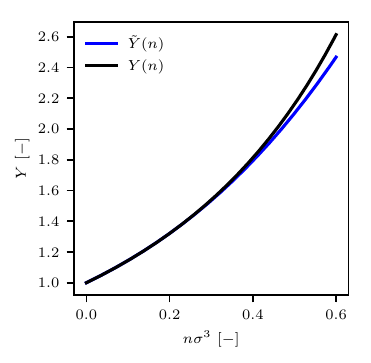}
    \caption{\color{ms4} Pair correlation function $Y(n)$ and polynomial approximation $\tilde{Y}(n)$ used in the Maxwell solution against normalized density for $n\sigma^3\in(0,0.6)$.}
    \label{fig1}
\end{figure}

\begin{figure}
  	\centering
	\begin{tabular}{cc}
  \includegraphics{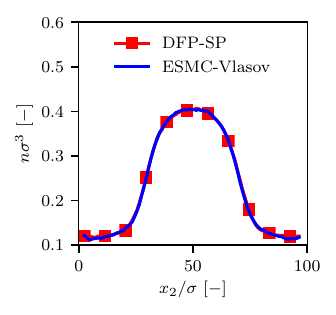}
  &\includegraphics{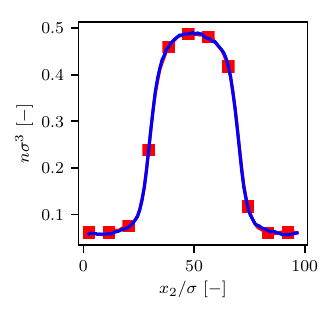} \\
  (a) $T = 180\ \mathrm{K}$& (b) $T = 170\  \mathrm{K}$\\
   \includegraphics{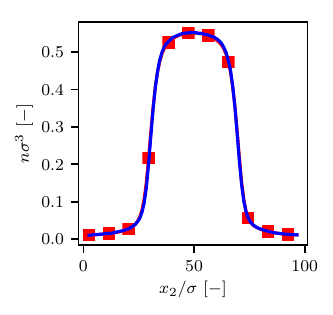}
  &\includegraphics{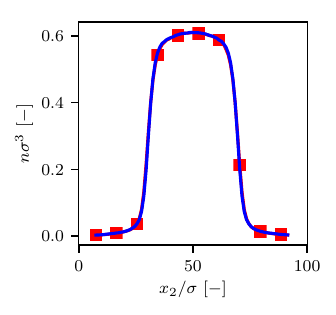} \\
  (c) $T = 160\  \mathrm{K}$ & (d)
  $T = 150\ \mathrm{K}$
  \end{tabular}
  \caption{\color{ms4} Number density profiles for simulation of DFP-SP model (in red) and ESMC-Vlasov solution algorithm  (in blue) for the obtained isothermal equilibrium simulation in $T=180, 170, 160\ \mathrm{and}\ 150 \ \mathrm{K}$ in (a)-(d), respectively.}
  \label{fig2}
\end{figure}

\begin{figure}
  	\centering
	\begin{tabular}{cc}
  \includegraphics{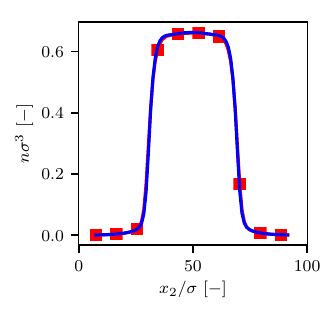}
  &\includegraphics{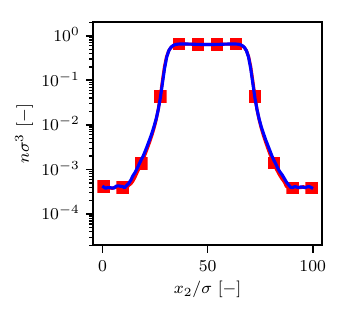} \\
  (a1) $T=140\ \mathrm{K}$& (a2) $T=140\ \mathrm{K}$, log-scale\\
  \includegraphics{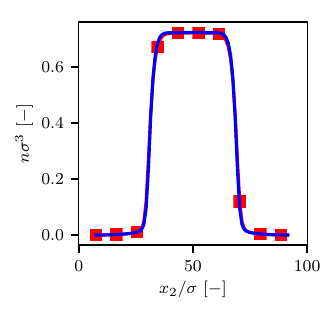}
  &\includegraphics{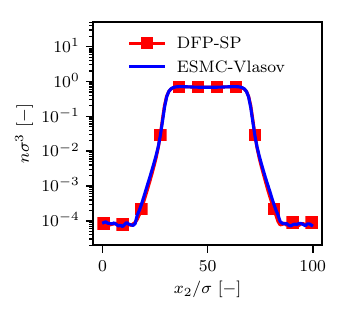} \\
  (b1) $T=130\ \mathrm{K}$& (b2) $T=130\ \mathrm{K}$, log-scale
  \end{tabular}
  \caption{\color{ms4} Number density profiles for simulation of DFP-SP model (in red) and ESMC-Vlasov solution algorithm  (in blue) for the obtained isothermal equilibrium simulation with $T=140, 130 \ \mathrm{K}$ in (a1-a2) and (b1-b2), respectively.}
  \label{fig3}
\end{figure}

\begin{figure}
\centering
	\begin{tabular}{cc}
\includegraphics{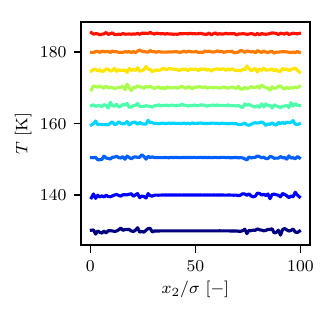}
  &\includegraphics{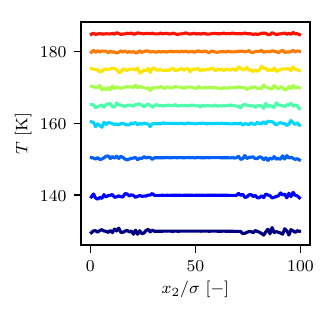} \\
  (a) DFP-SP & (b) ESMC-Vlasov
  \end{tabular}
  \caption{\color{ms4} Temperature profiles obtained from simulation of DFP-SP model (a) and ESMC-Vlasov solution algorithm  (b) for co-existing state test case at temperatures $T\in\{130, 140,150,160,165,170,175,180,185\} \ \mathrm{K}$ shown in rainbow spectrum, from dark blue to red.}
  \label{fig4}
\end{figure}

\begin{figure}
  	\centering
	\begin{tabular}{cc}
  \includegraphics{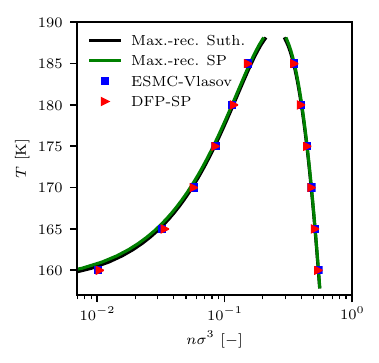}
  &\includegraphics{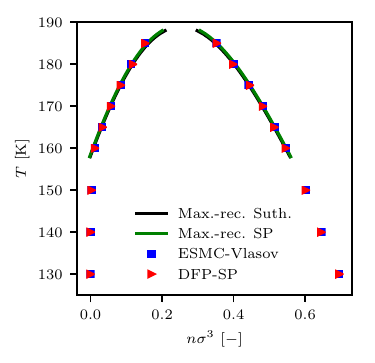} \\
  (a) co-existing states { for $n_l \sigma^3\leq 0.55$ in log-scale}& (b) co-existing states, extended
  \end{tabular}
  \caption{\color{ms4} Co-existing states  obtained  from simulations of DFP-SP model (in red), ESMC-Vlasov solution algorithm  (in blue), and Maxwell area reconstruction solution of Sutherland (in black) and Screened-Poisson potential (in green) for  equilibrium at $T\in\{130, 140,150,160,165,170,175,180,185\} \ \mathrm{K}$ with relative error in density of vapour  $|n_v^\mathrm{DFP-SP}-n_v^\mathrm{ESMC-Vlasov}|/n_v^\mathrm{ESMC-Vlasov}={0.052, 0.037,0.002,0.004,0.06,0.039,0.001,0.05,0.01}$, respectively. Density for the Monte Carlo solution is obtained by averaging over cells within $x_2\in (0,5\sigma)\cup(L-5\sigma,L)$ and $x_2\in (L/2-5\sigma, L/2+5\sigma)$ for vapour and liquid, respectively. }
  \label{fig5}
\end{figure}

}
\subsection{Evaporation near equilibrium}
\label{sec:evap_to_eq}
\begin{figure}
  \centering
  \includegraphics{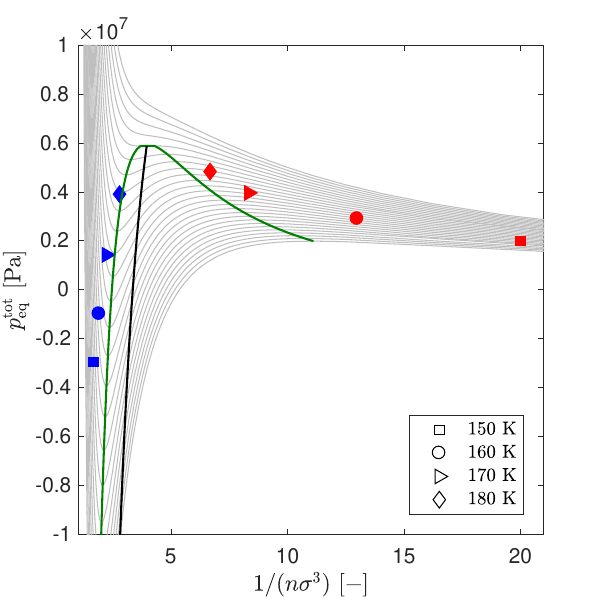}
  \caption{Contours of total equilibrium pressure $p_{\mathrm{eq}}^{\mathrm{tot}}(n,T){\color{ms3}:=nk_bT(1+nbY)+p^{\mathrm{att}}_{\mathrm{eq}}}$ at constant temperatures as well as the initial state pairs $(n_{\mathrm{liq}},T_{\mathrm{liq}})$ and $(n_{\mathrm{vap}},T_{\mathrm{vap}})$ for the evaporation test case with initial temperatures of $T^{(0)}\in\{ 150,160,170,180 \} \ \mathrm{K}$ and corresponding vapour (in red) and liquid (in blue) number densities of $n_{\mathrm{liq}}^{(0)}=\{0.6,0.526,0.45,0.36 \} \times \sigma^{-3} \ \mathrm{m}^{-3}$ and $n_{\mathrm{vap}}^{(0)}=\{ 0.05,0.077,0.12,0.15 \} \times  \sigma^{-3} \ \mathrm{m}^{-3}$, respectively. {\color{ms3}Note that  here the total pressure with attractive part associated with {\color{ms4}Screen-Poisson}'s potential is used $p^\mathrm{att}_{\mathrm{eq}}(n)={\color{ms4}  {a }  { e^{-\lambda \sigma} }{  }(\lambda^2 \sigma^2 + 3 \sigma \lambda +3 ) n^2/(6 \lambda^2)}$.}}
  \label{fig6}
\end{figure}

\begin{figure}
  	\centering
	\begin{tabular}{cc}
  \includegraphics{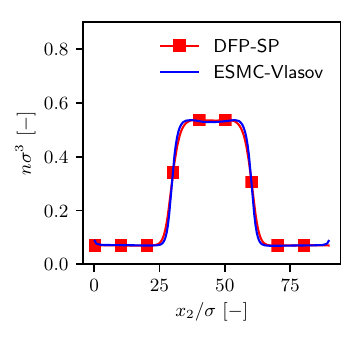}
  \
  &\includegraphics{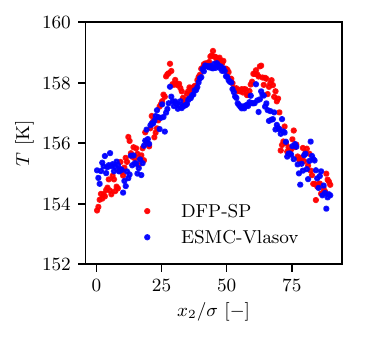} \\
  (a) & (b) \\
  \includegraphics{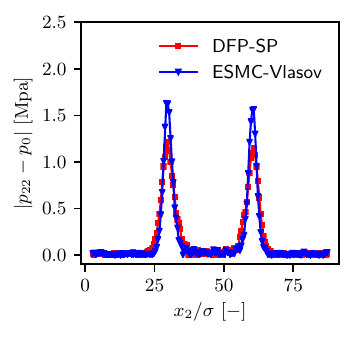}
  \
  &\includegraphics{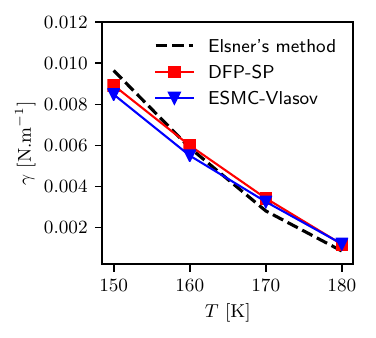} \\
  (c) & (d)
  \end{tabular}
  \caption{Phase transition of Argon at $T^{(0)}=160\ \mathrm{K}$. (a) Normalized number density. (b) Normalized temperature. (c) Normalized difference between diagonal and equilibrium kinetic pressure. (d) Normalized surface tension for liquid temperatures of $T^{(0)}\in \{ 150,160, 170, 180 \}\ \mathrm{K}$.{ \color{ms3} Here, maximum relative error of \%7 in evaluating $\gamma$ is observed.}}
  \label{fig7}
\end{figure}

\begin{figure}
  \centering
  \includegraphics{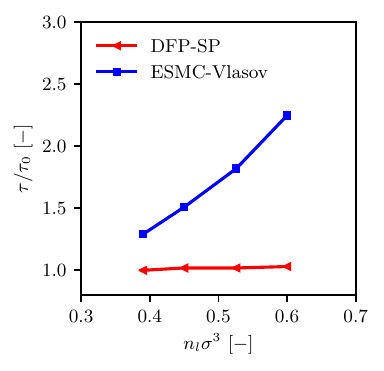}
  \caption{Execution time per particle $\tau:=t^{\mathrm{ex.}}/N_p$ for evaporation simulation at the equilibrium using ESMC-Vlasov and DFP-SP solution algorithms. The cost is normalized by $\tau_0$ denoting DFP-SP execution time for the case $T^{(0)}=180$~K.}
  \label{fig8}
\end{figure}

\noindent In order to present DFP-SP prediction of the surface tension, evaporation of a liquid Argon slab in contact with the vapor phase {\color{ms4} near} equilibrium is studied. In particular, we consider the phase transition of the liquid slab located within  $[0,1]\times [0,30 \sigma] \times [0,1]\ m^3$ which is surrounded by the vapour confined in  $[0,1]\times [0,L] \times [0,1]\ m^3$ with $L=90 \sigma$. Since only $x_2$ direction is relevant, the other dimensions in the physical space are ignored in the evolution of particles positions. The liquid slab and the vapour around it, are initialized with number densities $n_{\mathrm{liq}}$ and $n_{\mathrm{vap}}$, respectively. In order to maintain the stationary scenario, the liquid slab is heated up every $N_{\mathrm{therm}}$ steps, where the particles inside   $x_2\in[L/2-L_{\mathrm{therm}}/2,L/2+L_{\mathrm{therm}}/2]$ are thermostated with the initial temperature $T^{(0)}$. This is achieved by re-scaling particles fluctuating velocities.  Furthermore, the specular reflection boundary conditions are deployed here.
\\ \ \\
After conducting convergence studies, it is found that $N_{\mathrm{cell}}=180$ computational cells and the CFL number $0.01$ provide converged results.
Moreover, a fixed statistical weight of  $w=10^{14}$ was picked
for the test cases with $T^{(0)} \in \{150, 160, 170, 180 \}\ \mathrm{K}$, $n_{\mathrm{liq}}^{(0)}\in\{0.6,0.526,0.45,0.36 \} \times \sigma^{-3} \  \mathrm{m}^{-3}$ and $n_{\mathrm{vap}}^{(0)} \in \{ 0.05,0.077,0.12,0.15 \} \times  \sigma^{-3} \ \mathrm{m}^{-3}$. The stationary distribution is obtained after $5'000$ steps and the time averaging is performed over  $15'000$ steps onward. Here, $L_{\mathrm{therm}}=10 \sigma$ was chosen as the length of thermostated sub-domain which was carried out once every $N_{\mathrm{therm}}=1'000$ steps. The initial densities of vapour and liquids for each simulation were picked such that unstable regions of the phase diagram are not met, as depicted in Fig.~\ref{fig6}. {\color{ms2} Here, total equilibrium pressure is $p_{\mathrm{eq}}^{\mathrm{tot}}(n,T) :=nk_bT(1+nbY)+p^{\mathrm{att}}$, where the attractive part 
\begin{flalign}
p^\mathrm{att}_{\mathrm{eq}}(n)=-({2\pi}/{3})n^2 \int_0^\infty r^3 \dfrac{\partial \phi(r)}{\partial r}  dr
\end{flalign}
  becomes  $p^\mathrm{att}_{\mathrm{eq}}(n)=-({4 \pi }/{3}) \phi_0 \sigma^3 n^2$ and $p^{\text{att}}_{\mathrm{eq}}  (n) =    {a }  { e^{-\lambda \sigma} }{  }(\lambda^2 \sigma^2 + 3 \sigma \lambda +3 ) n^2/(6 \lambda^2)$ in case of Sutherland and Screened-Poisson potentials, respectively \cite{sadr2019treatment}. 
}
\\ \ \\
For the liquid slab at equilibrium, the surface tension coefficient $\gamma$ is defined as a norm of the difference between diagonal element of the kinetic pressure tensor and the equilibrium kinetic pressure across the phase transition
\begin{flalign}
\gamma = \frac{1}{2} \int_0^L (p_{22}-p_0) dx_2~.
\end{flalign}
For details see \cite{ghoufi2010calculation,Hirschfelder1963}. The integration over the domain can be taken numerically, since the discrete values of the pressure tensor are estimated at every cell. 
\\ \ \\
Comparison between simulation results obtained from the DFP-SP model with respect to the benchmark ESMC-Vlasov is shown in Fig.~\ref{fig7}. 
Although the DFP-SP model under-predicts the peak in estimating $|p_{22}-p_0|$ at the interface, overall a very good agreement between the two is observed. Note that the discrepancy in $\gamma$ between both methods increases as the liquid gets denser at lower temperatures. 
Furthermore, the computational costs of the discussed simulation settings are presented in Fig~\ref{fig8}.  {\color{ms3} Accurate quantification of the relative computational costs would require a well-designed prospective study with detailed numerical investigations. However a general comparison is possible by comparing the complexity order of ESMC-Vlasov and DFP-SP methods. For the former, observe that the number of operations follows $\mathcal{O}(n\sqrt{T})$.} Yet the cost of the latter is independent of the density, { \color{ms3}which makes our assembled approach attractive for efficient Monte-Carlo simulations of multi-phase flows}.
{
\color{ms2}
\\ \ \\
Next, we compare the obtained numerical results in estimating the surface tension coefficient with respect to an analytical approximation based on the free surface energy $S(T)$. Following \cite{elsner1991calculation}
\begin{flalign}
\gamma(T) = T \int_T^{T_\mathrm{crit}} \frac{S(T)}{T^2} dT,
\end{flalign}
where
\begin{flalign}
S(T) = \frac{\alpha}{3} \int_{\vartheta _v}^{\vartheta _l} (\frac{\mu}{\vartheta }-p_v)\frac{d\vartheta }{\vartheta ^{2/3}}.
\end{flalign}
Here, $\vartheta =1/n$ denotes the reciprocal number density, $\mu$ is the chemical potential, $p_v$ is the vapour pressure and $\alpha$ a proportionality factor. Note that $\mu-p_v \vartheta $  gives the Helmholtz free energy per particle.
For simplicity, similar to \cite{hadjiconstantinou2000surface}, we set $\alpha=0.6$ (for all temperatures). Furthermore, we calculate the chemical potential through
\begin{flalign}
\frac{\mu}{T} &= \frac{\mu_\mathrm{crit.}}{T_\mathrm{crit.}} - \int_{p_v/T}^{p_v/T} \vartheta  d(p_v/T),
\end{flalign}
where
\begin{flalign}
\frac{\mu_\mathrm{crit.}}{T_\mathrm{crit.}} &= - k_b (2\xi_\mathrm{crit.} + \ln(\vartheta _\mathrm{crit.}/\vartheta _{q,\mathrm{crit.}})),\\
\xi_c &= \frac{\vartheta _\mathrm{crit.} p_\mathrm{crit.}}{k_b T_\mathrm{crit.}}\\
\textrm{and} \ \ \ \ \vartheta _{q} &= \frac{\hbar^3}{(2\pi m k_b T)^{3/2}}.
\end{flalign}
Note that $\hbar$ is the Planck constant (see \cite{elsner1989interaction} for details). Figure~\ref{fig7}-d shows reasonable agreement between all three considered methods i.e. analytical Elsner's approach, the DFP-SP model and the ESMC-Vlasov simulation results.
}


\subsection{Evaporation to vacuum}
\begin{figure}
  \centering
  	\begin{tabular}{cc}

  \includegraphics{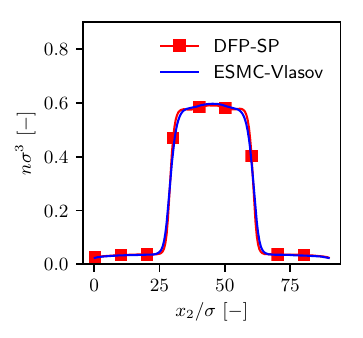}
  \
  &\includegraphics{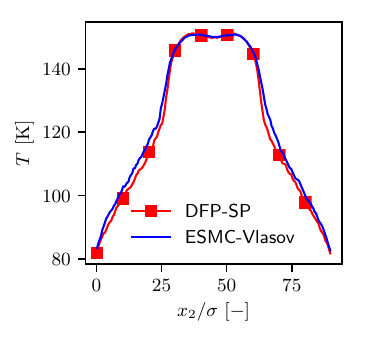} \\
  (a) & (b) \\
  \includegraphics{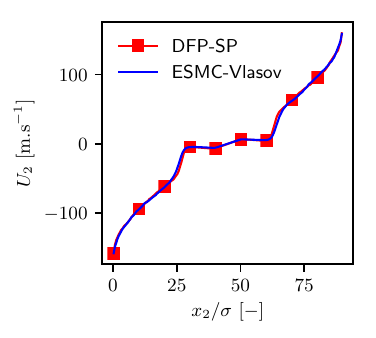}
  \
  &\includegraphics{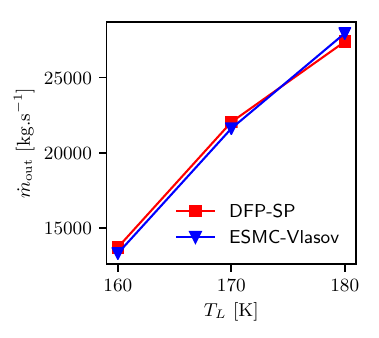} \\
  (c) & (d)
  \end{tabular}
  \caption{Simulation results for evaporation of liquid Argon into the vacuum obtained from ESMC-Vlasov and DFP-SP models. (a) Normalized number density. (b) Temperature. (c) Mean velocity. (d) Outflow mass flux.}
  \label{fig9}
\end{figure}

\noindent Since streaming of the stochastic particle in the DFP model is consistent with the Enskog equation only on the moment level, the measurement of evaporation rate based on particle trajectory as suggested in \cite{yasuoka1994evaporation,kon2014method,kobayashi2016molecular} would not lead to an accurate estimation of the desired property. Alternatively, the evaporation rates can be approximated via moments as a liquid slab evaporates into the vacuum, see \cite{frezzotti2005mean}. The outflow mass flux $\dot{m}_{\text{out}}$ is indeed a function of the evaporation rate $\sigma_e$
\begin{flalign}
\dot{m}_{\text{out}} 
=
A \sigma_e n_g(T_L) \sqrt{\frac{k_b T_L}{2 \pi m}},
\end{flalign}
\noindent where $A$ indicates the area of cross-section normal to the inter-phase, $n_g(T_L)$ is the number density of the gas computed at the temperature of the liquid phase $T_L$. Hence, validation for the evaporation rate translates into the  consistency of temperature and density profiles, as well as the outgoing mass flux with respect to the benchmark solution. Except for the boundary conditions, we consider similar simulation settings as the test case with the evaporation of liquid at equilibrium, i.e., \S~\ref{sec:evap_to_eq}. Particles are re-initialized in the liquid slab as they leave the domain with the new position sampled randomly from a uniform distribution on the interval $x_2\in[L/2-L_{\mathrm{therm}}/2,L/2+L_{\mathrm{therm}}/2]$. For the re-initialized particle, the velocity component in the relevant flow direction, i.e., $V_2$, is sampled again from an equilibrium distribution with mean and temperature of the local computational cell, as the other components of the velocity remain unchanged. 
\\ \ \\
In comparison to ESMC-Vlasov, we obtain accurate predictions of  temperature, density, velocity and the outflow mass flux, as shown in Fig.~\ref{fig9}. Similar test cases were performed at other temperatures and again good agreements between the DFP-SP model and the benchmark were observed, indicating consistent evaporation rates.
\section{Inverted temperature gradient}
\label{sec:inverted}
\begin{figure}
  \centering
  \includegraphics{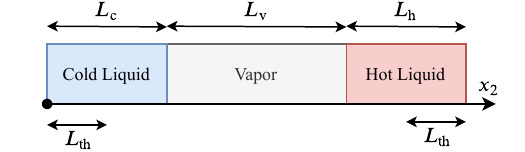}
  \caption{Schematic of the inverted temperature gradient simulation}
  \label{fig10}
\end{figure}

\noindent One of the interesting examples where counter-intuitive non-equilibrium effects become evident is the inverted temperature profile in the vapor phase contained between two liquid slabs at different temperatures. First, \cite{pao1971application} found out that the temperature profile does not increase monotonically from cold to hot liquid, suggesting an inverted temperature profile in the vapor phase. Since then, this phenomenon has been studied and validated in details using tools across Molecular Dynamics and  kinetic theory (see e.g. \cite{frezzotti2003evidence,meland2003molecular,meland2004dependence}). 
\\ \ \\
We tackle this problem by the devised DFP-SP model, and validate our results with respect to the ESMC-Vlasov approach. Consider a solution domain in the physical space $\Omega=[0,1]\times[0,L]\times [0,1]\ \mathrm{m}^3$ where $L=70 \sigma$. Cold and hot liquid slabs are located at $x_2\in[0,L_c]$ and $x_2\in[L-L_h,L]$, respectively, with $L_c=L_h=15 \sigma$. The thermodynamic states of cold and hot slabs are initially set by the density-temperature pairs $(n_c,T_c)$ and $(n_h,T_h)$, respectively. The vapour with initial density-temperature of $(n_v,T_v)$ fills the space between two slabs as depicted schematically in Fig.~\ref{fig10}.  In order to keep temperature of the liquid slabs close to the initial setting, particles within $x_2\in [0,L_{\mathrm{therm}}]$ and $x_2\in [L-L_{\mathrm{therm}}, L]$, where $L_{\mathrm{therm}}=15 \sigma$, are thermostated to the initial temperature every $N_{\mathrm{therm}}$ steps. This is done by re-scaling particles fluctuating velocities. Due to symmetry of the simulation setting, position evolution of the particles is considered only along the $x_2$-coordinate. 
\\ \ \\
Fully diffusive scattering is applied as the boundary condition to the both ends of the domain. As particles leave the domain, their velocities are sampled from the flux of the Maxwellian distribution with the initial temperature of the nearest liquid slab. Next the scattered particles are streamed into the domain for the remaining fraction of the time step.
\\ \ \\
Since there exists a mass flux from hot slab towards to the cold one, the system does not admit a stationary state. Hence time averaging can not be deployed to improve the estimate of observables. However, by moving the observer in the direction opposite to the mass flux, a quasi-stationary condition is reached where time averaging becomes applicable. Notice that a similar approach is typically utilized for one-dimensional shock wave problems \cite{Bird,gorji2013fokker}.  Once every $N_{\mathrm{shift}}$ steps, the simulation frame is moved by $\Delta/2$ where $\Delta:=(\mathcal{V}_c-\mathcal{V}_h)$, $\mathcal{V}_c:=x_L$ and  $\mathcal{V}_h:=(L-x_R)$. Here $x_R$ is the right-most location with number density $0.9n_h$, whereas $x_L$ is the left-most position with number density $0.9n_c$. As the simulation frame is moved, the particles which appear outside of the domain are removed. Moreover, new particles are generated to fill the void space. Velocities of the newly introduced ones are sampled from the Maxwellian distribution with temperature of the nearest liquid slab. Their corresponding positions are sampled uniformly inside the void space.
\\ \ \\
Three settings with temperatures below the critical point are considered as presented in Table~\ref{tab:inverted_initial}. The temperature difference between liquid slabs is set to $2\ \mathrm{K}$. After a convergence study, a uniform discretization of physical space with mesh size of $\Delta x_2=\sigma/2$ along $x_2$ direction as well as the CFL number $0.01$ are picked due to their reasonable accuracy. In all simulations, initially $10,000$ particles per cell for the vapour phase are generated, leading to the statistical weights $w \in \{4.3, 5.1,5.6 \} \times 10^{13}$ associated with cold liquid slab temperatures $T_{c} \in \{ 165,167,170 \}\  \mathrm{K}$, respectively. Furthermore, the domain is shifted every $N_{\mathrm{shift}}=500$ steps and the liquid slabs are thermostated every $N_{\mathrm{therm}}=1'000$  steps, for all considered cases.
\\ \ \\
\begin{table}
  \begin{center}
\def~{\hphantom{0}}
  \begin{tabular}{cccccc}
      Test Case & $n_c \sigma^3\ [-]$ & $n_h \sigma^3\ [-]$ &$n_v \sigma^3\ [-]$ &$T_c\ [\mathrm{K}]$ & $T_h\ [\mathrm{K}]$ \\[3pt]
       (\RNum{1})   & $0.42$ & $0.40$ & $0.13$ &  $170$ & $172$\\
       (\RNum{2})   & $0.45$ &  $0.43$ & $0.12$ & $167$ & $169$\\
       (\RNum{3})  & $0.47$ & $0.45$ & $0.10$ & $165$  & $167$\\
  \end{tabular}
  \caption{Initial number density and temperature of liquid slabs and the vapour for the inverted temperature gradient simulations}
  \label{tab:inverted_initial}
  \end{center}
\end{table}

\begin{figure}
  \centering
  \begin{subfigure}[b]{0.31\columnwidth}
   \scalebox{0.9}{
  \includegraphics{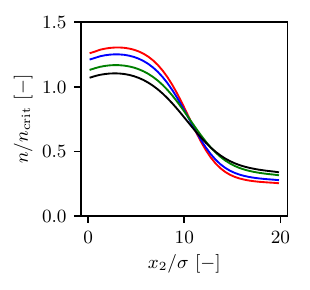}
  }
  \caption{Number density}
  \end{subfigure}
   \  \
  \begin{subfigure}[b]{0.31\columnwidth}
   \scalebox{0.9}{
  \includegraphics{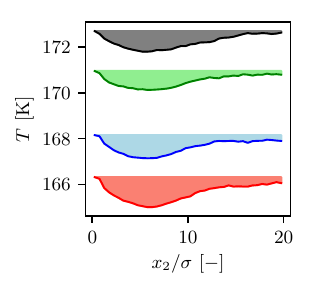}
  }
  \caption{Temperature}
  \end{subfigure}
  \ \ 
  \begin{subfigure}[b]{0.31\columnwidth}
  \scalebox{0.9}{
   \includegraphics{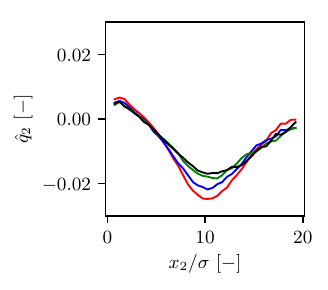}}
   \caption{Heat-flux}
   \end{subfigure}
   \caption{Single droplet evaporation. The profile of number density, temperature and heat-flux for the liquid/vapour contained between diffusive and specular walls.}
   \label{fig11}
\end{figure}

\begin{figure}
  \centering
  \begin{subfigure}[b]{0.31\columnwidth}
  \includegraphics{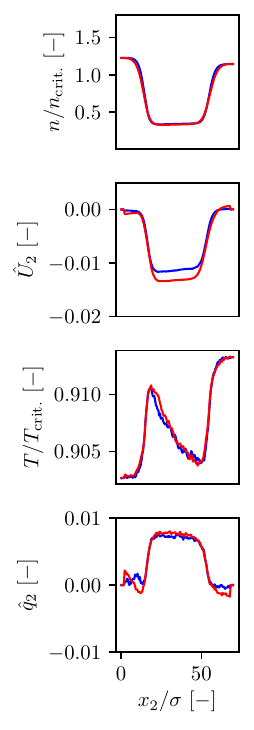}
  \caption{Test case (\RNum{1})}
  \end{subfigure}
  \begin{subfigure}[b]{0.3\columnwidth}
  \includegraphics{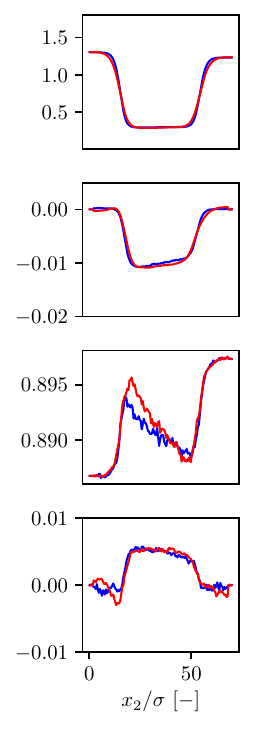}
  \caption{Test case (\RNum{2})}
  \end{subfigure}
  \begin{subfigure}[b]{0.3\columnwidth}
   \includegraphics{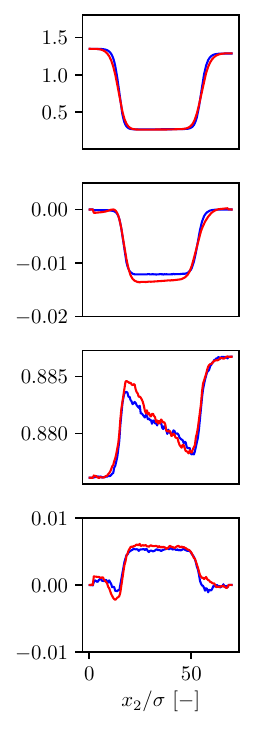}
   \caption{Test case (\RNum{3})}
   \end{subfigure}
   \caption{Normalized number density, mean velocity, temperature, and heat-flux obtained from inverted temperature gradient simulations using ESMC-Vlasov (blue) and DFP-SP (red) solution algorithms. 
The critical point  $(n_{\mathrm{crit.}},T_{\mathrm{crit.}})=(0.3568/\sigma^3 \ \mathrm{m}^{-3},188.33\ \mathrm{K})$
 is used to normalize density and temperature. Heat-flux and mean velocity are normalized by $\hat{q}_2:=q_2/\left(\rho_c (kT_c/m)^{3/2}\right)$ and $\hat{U}_2:=U_2/\sqrt{kT_c/m}$, respectively.}
   \label{fig12}
\end{figure}

\noindent As depicted in Fig.~\ref{fig12}, the DFP-SP solution algorithm approximates the density, the mean velocity, the temperature and the heat-flux of the considered test cases accurately in comparison to the benchmark. The inverted temperature gradients are well recovered by the presented DFP-SP model. Furthermore, the profiles near the phase-transition regimes are found to be well approximated. 
As suggested by \cite{meland2004dependence}, since the colder liquid slab has smaller evaporation and condensation rates, we observe an accumulation of hot particles near the colder liquid slab. 
This leads to the conclusion that particles emitted from the hot liquid slab plays a more important role behind this counter intuitive non-equilibrium temperature profile.  
\\ \ \\
In order to obtain a better understanding of the underlying phase-transition dynamics, let us consider the case where a single liquid slab distributed in $x_2 \in [0,10\sigma]$ is in contact with a diffusive wall at $x_2=0$ and is surrounded by the vapour distributed inside $x_2 \in [10\sigma,20\sigma]$. For simplicity, we consider the specular wall at the end of domain $x_2=20 \sigma$ to contain the vapor. We simulated this setting using the DFP-SP model for liquid temperatures considered in the inverted-temperature-gradient test case using $\omega=4\times 10^{13}$, CFL number $0.01$ and $\Delta x = \sigma/2$. As shown in Fig.~\ref{fig11}, the temperature of the fluid first drops during the transition of the liquid to the vapor and then raises back inside the vapor phase. It is important to notice that the further we move far from the droplet, the more increase we observe in the temperature. This can be explained by noting that at relatively low vapor pressure, high energy particles can travel longer distances before reaching equilibrium with the background gas. Here, one can make three observations. First, note that as the temperature of the liquid decreases, the difference between temperature of the liquid and its surrounding vapour becomes larger. Next, the temperature drop gets more pronounced, once colder liquid slabs are considered. Finally, by a simple super-position of the observed single liquid slab temperature profiles, we get a qualitative picture of the inverted temperature gradient behaviour consistent with our two liquid slabs system shown in Fig.~\ref{fig12}.

\section{Spinodal decomposition}
\label{sec:spinodal_decomp}
\begin{figure}
\begin{tikzpicture}
\begin{scope}
    \node[anchor=south west,inner sep=0] (image) at (0,0) {\scalebox{0.85}{\includegraphics{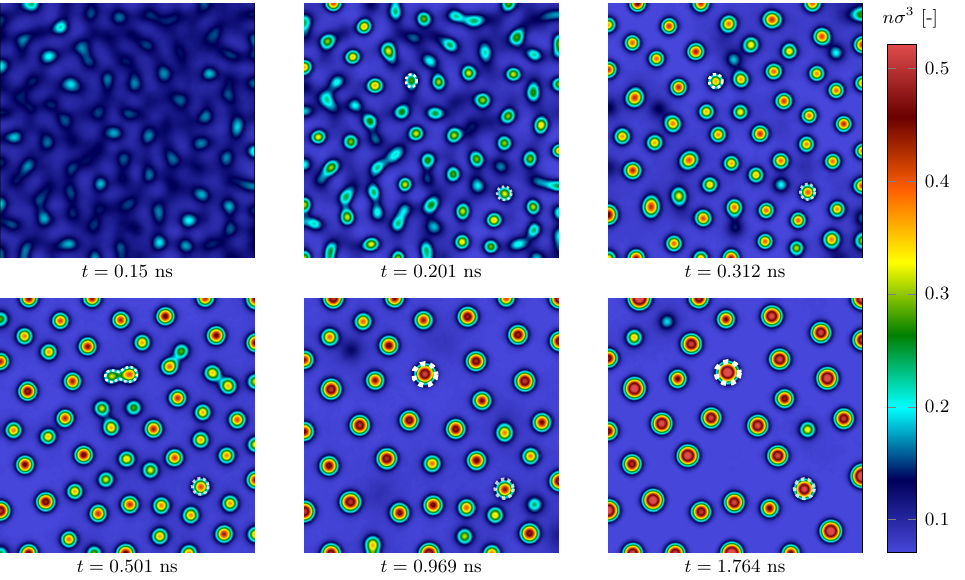}}};
    \begin{scope}[x={(image.south east)},y={(image.north west)}]
    \draw [-stealth, line width=1pt, cyan,anchor=west] (-0.01,0.35) -- ++(0.117,0.0) node (2)
[pos=-0.2]
{\footnotesize 2};
        \draw [-stealth, line width=1pt, red, anchor=west] (-0.01,0.16) -- ++(0.21,0.0) node (1) 
        [pos=-0.1]
        {\footnotesize 1};
    \end{scope}
\end{scope}
\end{tikzpicture}%
\caption{
Evolution of normalized number density in the spinodal decomposition of Argon obtained from simulation results of the DFP-SP model for test case (A), with initial number density $n^{(0)}=3\times10^{27}\ \mathrm{m}^{-3}$ and temperature $T^{(0)}=\ 120\ \mathrm{K}$ in the domain $\Omega_x=[0,L]^2\times[0,1]\ \mathrm{m}^3$ where $L=5\times10^{-8}\ \mathrm{m}$. Here, two droplets are labeled with rings around them which are tracked in time for further analysis.
}
   \label{fig13}
 \end{figure}

\begin{figure}
  \centering
  \begin{subfigure}[b]{0.49\columnwidth}
   \scalebox{0.37}{
  \includegraphics{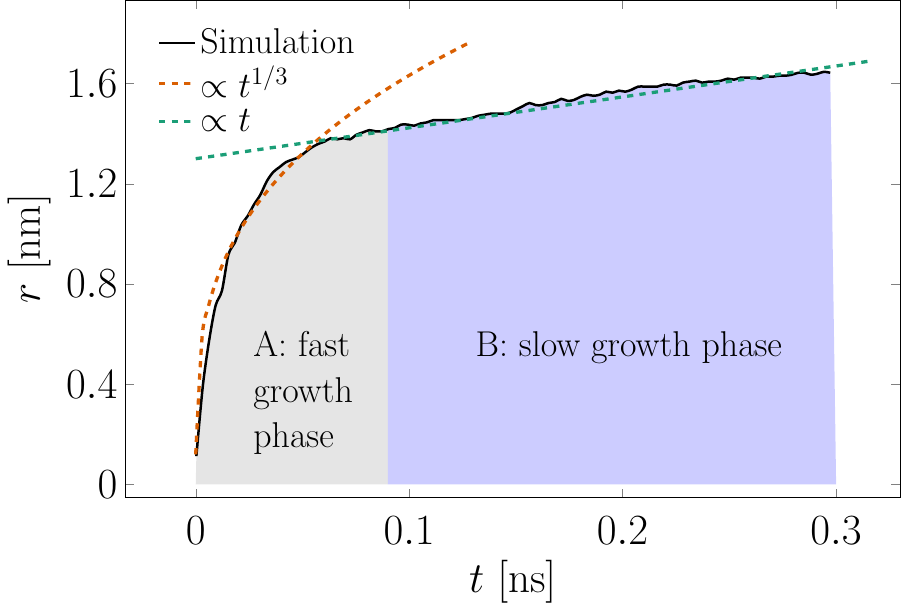}  }
  \caption{Droplet \#1}
  \end{subfigure}
   \hfill
  \begin{subfigure}[b]{0.49\columnwidth}
   \scalebox{0.37}{
  \includegraphics{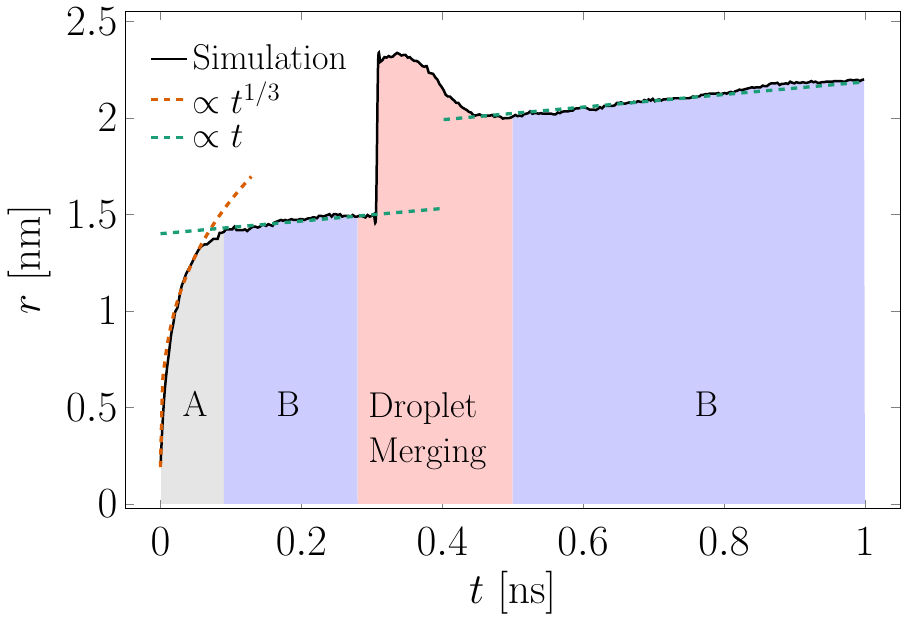}}
  \caption{Droplet \#2}
  \end{subfigure}
   \caption{
   Growth of the labeled droplets, i.e., droplet \#1 in (a) and \#2 in (b) as identified in Fig. \ref{fig13}, are studied by investigating the evolution of droplet's radius $r$ in time. We observe that initial growth of radius in both cases is $\propto t^{1/3}$ which follows by $\propto t$ unless interrupted by a coalescence, e.g. in (b).
   }
   \label{fig14}
\end{figure}

\begin{figure}
  \centering
  \begin{subfigure}[b]{0.44\columnwidth}
   \scalebox{0.37}{
  \includegraphics{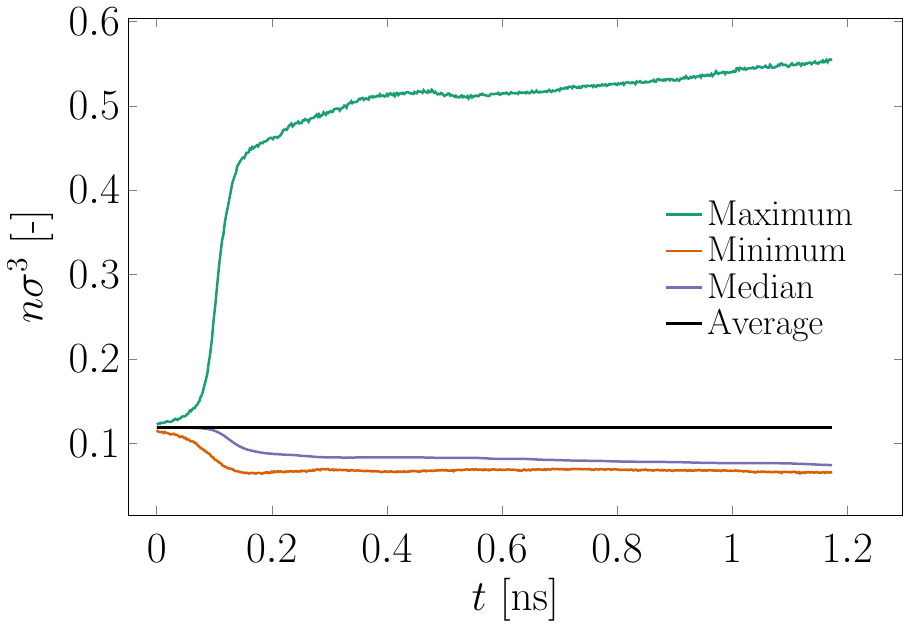}  }
  \caption{ }
  \end{subfigure}
   \hfill
  \begin{subfigure}[b]{0.44\columnwidth}
   \scalebox{0.37}{
  \includegraphics{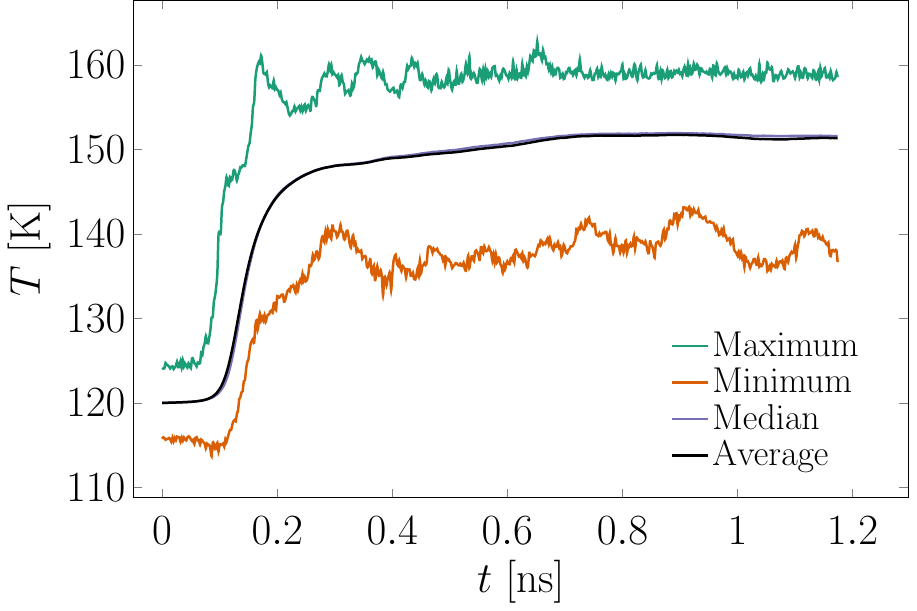}}
  \caption{ }
  \end{subfigure}
   \caption{Evolution of maximum, minimum, median and average values of normalized number density as well as temperature in the whole simulation domain for the test case (A). Results are obtained using the DFP-SP model.}
   \label{fig15}
\end{figure}

\begin{figure}
  \centering
  \begin{subfigure}[b]{0.44\columnwidth}
   \scalebox{0.37}{
  \includegraphics{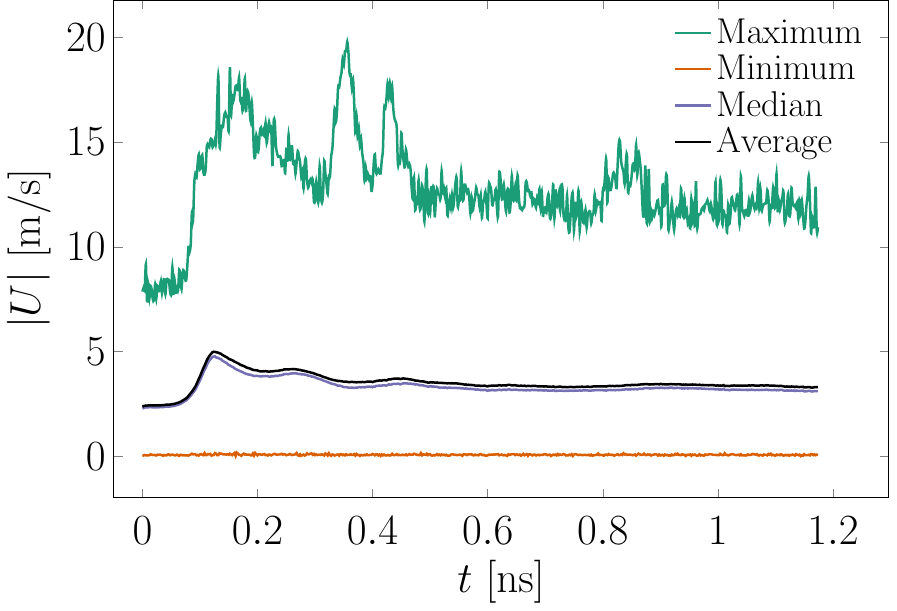}  }
  \caption{ }
    \label{fig16a}
  \end{subfigure}
   \qquad
  \begin{subfigure}[b]{0.44\columnwidth}
   \scalebox{0.36}{
  \includegraphics{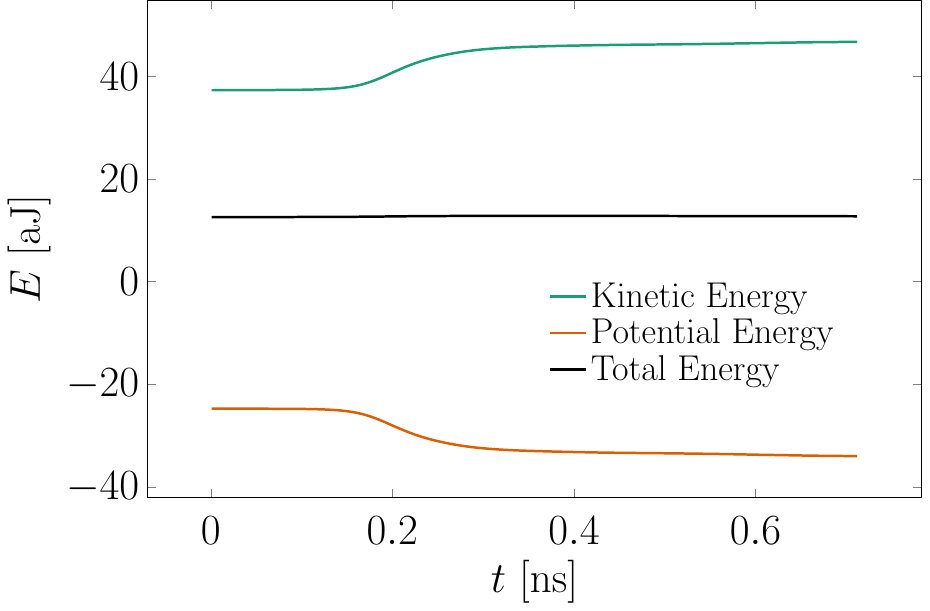}}
  \caption{ }
  \label{fig16b}
  \end{subfigure}
   \caption{
   Evolution of maximum, minimum, median and average values of velocity magnitude  in the whole simulation domain for the test case (A) shown in (a). The evolution of kinetic, potential and total energy of the system are depicted in (b). Here, $[\mathrm{a}]$ denotes the prefix unit called atto where $\mathrm{a}=10^{-18}$. Results are obtained using the DFP-SP model.
   }
   \label{fig16}
\end{figure}

\begin{figure}
  \centering
   \scalebox{0.85}{
  \includegraphics{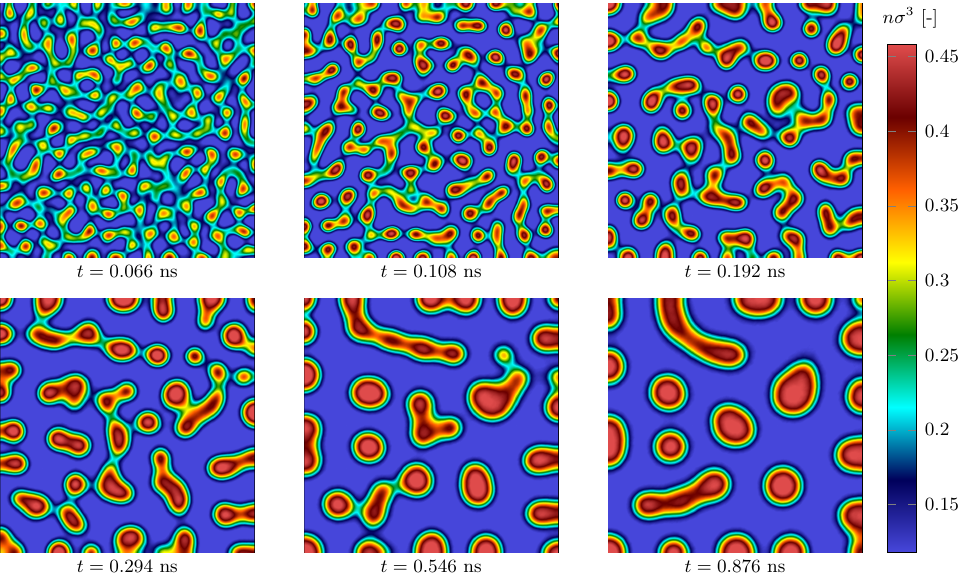}}
   \caption{
   Evolution of normalized number density in the spinodal decomposition of Argon obtained from simulation results of the DFP-SP model  for test case (B),  with initial number density $n^{(0)}=5\times10^{27}\ \mathrm{m}^{-3}$ and temperature $T^{(0)}=\ 120\ \mathrm{K}$ in the domain $\Omega_x=[0,L]^2\times[0,1]\ \mathrm{m}^3$ where $L=5\times10^{-8}\ \mathrm{m}$.
   }
   \label{fig17}
\end{figure}

\begin{figure}
  \centering
  \begin{subfigure}[b]{0.44\columnwidth}
   \scalebox{0.37}{
  \includegraphics{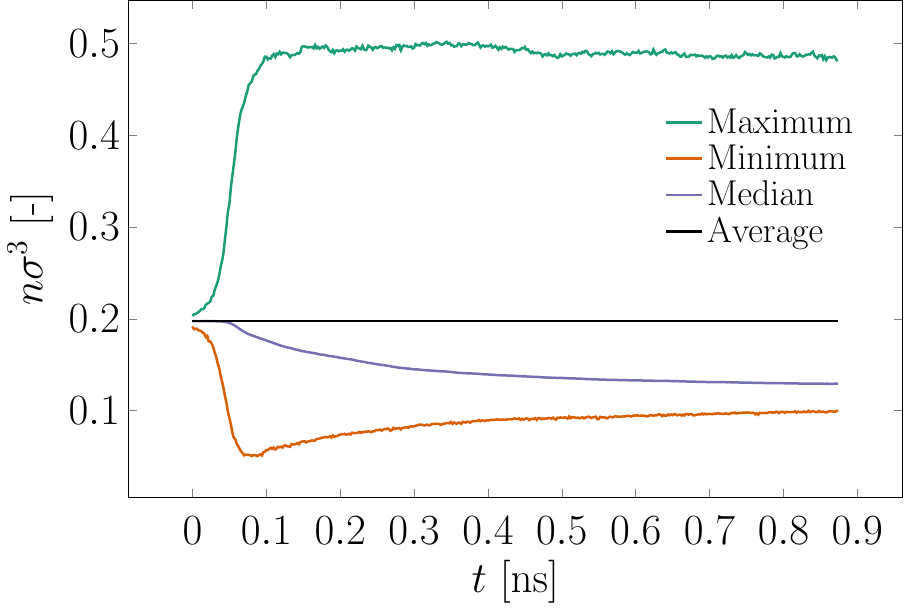} }
  \caption{}
  \end{subfigure}
   \hfill
  \begin{subfigure}[b]{0.44\columnwidth}
   \scalebox{0.37}{
  \includegraphics{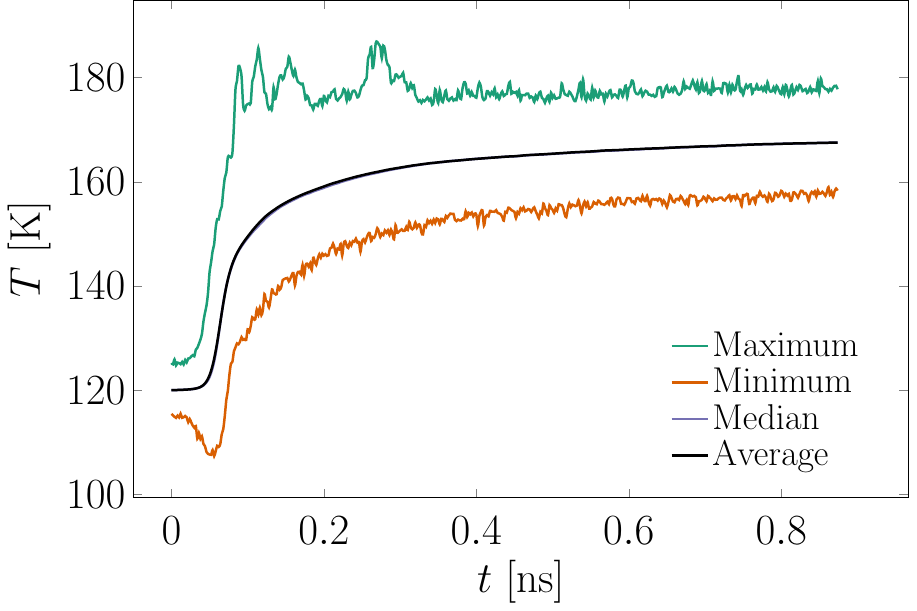}}
  \caption{}
  \end{subfigure}
   \caption{
   Evolution of maximum, minimum, median and average values of normalized number density as well as temperature in the whole simulation domain $\Omega_x$ for spinodal decomposition of Argon with settings of case (B). Results are obtained using the DFP-SP model.
   }
   \label{fig18}
\end{figure}

\begin{figure}
  \centering
  \begin{subfigure}[b]{0.44\columnwidth}
   \scalebox{0.37}{
 \includegraphics{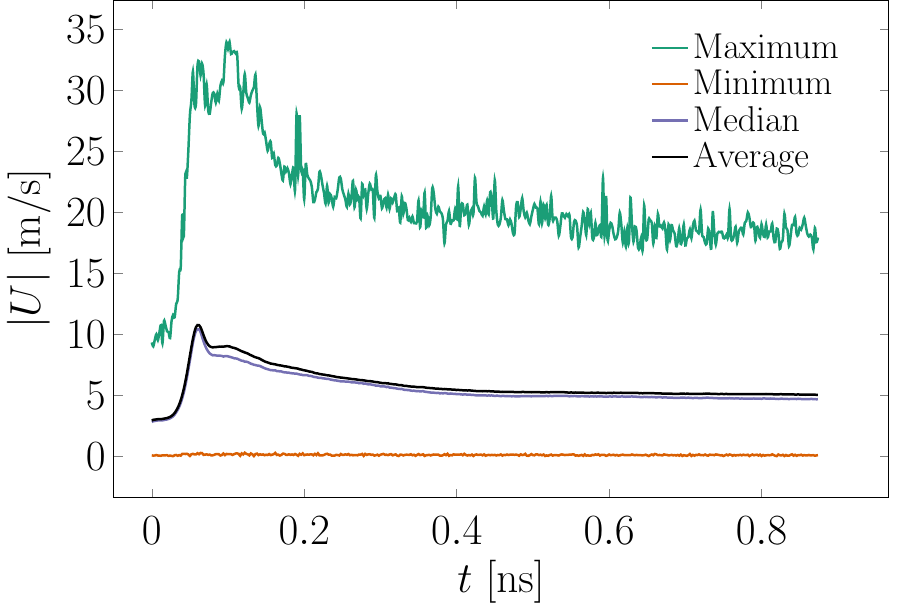}  }
  \caption{}
    \label{fig:meddensvelo}
  \end{subfigure}
   \qquad
  \begin{subfigure}[b]{0.44\columnwidth}
   \scalebox{0.36}{
  \includegraphics{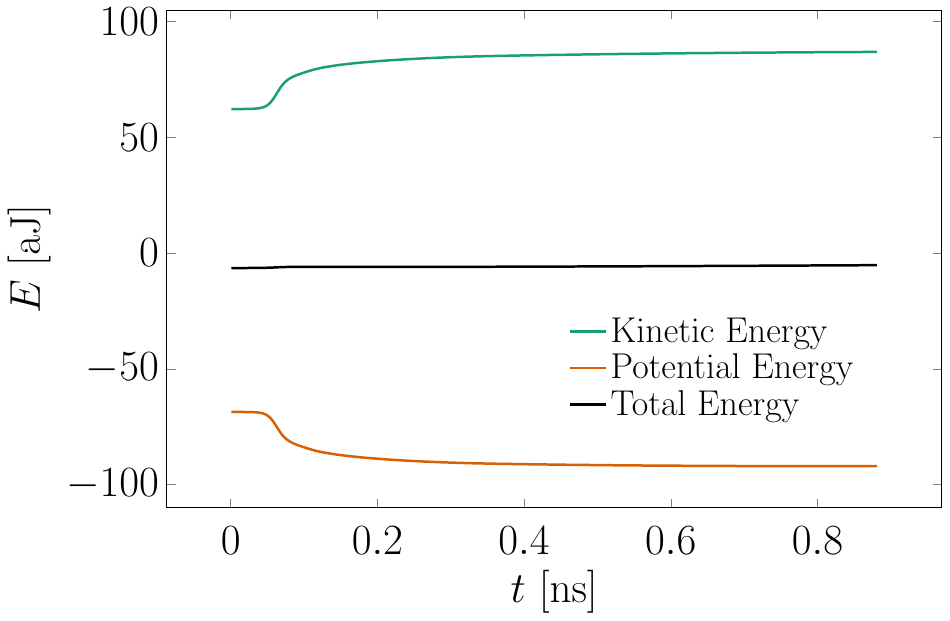}}
  \caption{}
  \label{fig19b}
  \end{subfigure}
   \caption{
    Evolution of maximum, minimum, median and average values of velocity magnitude  in the whole domain $\Omega_x$ for the test case (B) in simulating the spinodal decomposition using the DFP-SP model are shown in (a). The evolution of kinetic, potential and total energy of the system are depicted in (b).
   }
   \label{fig19}
\end{figure}

\begin{figure}
   \scalebox{0.85}{
  \includegraphics{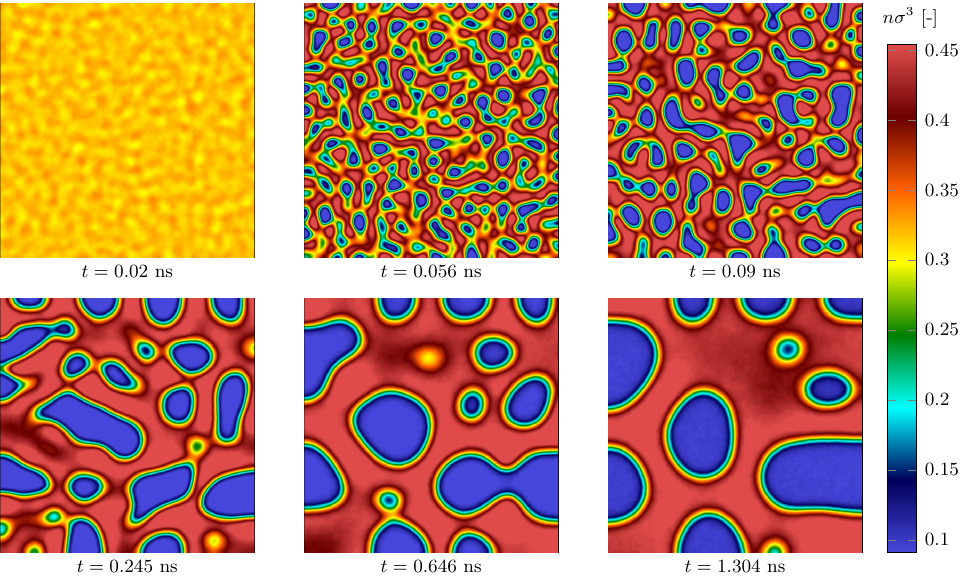}}
   \caption{
   Evolution of normalized number density in the spinodal decomposition of Argon obtained from simulation results of the DFP-SP model for test case (C),  with initial number density $n^{(0)}=8\times10^{27}\ \mathrm{m}^{-3}$ and temperature $T^{(0)}=\ 120\ \mathrm{K}$ in the domain $\Omega_x=[0,L]^2\times[0,1]\ \mathrm{m}^3$ where $L=5\times10^{-8}\ \mathrm{m}$.
   }
   \label{fig20}
\end{figure}

\begin{figure}
  \centering
  \begin{subfigure}[b]{0.44\columnwidth}
   \scalebox{0.37}{
  \includegraphics{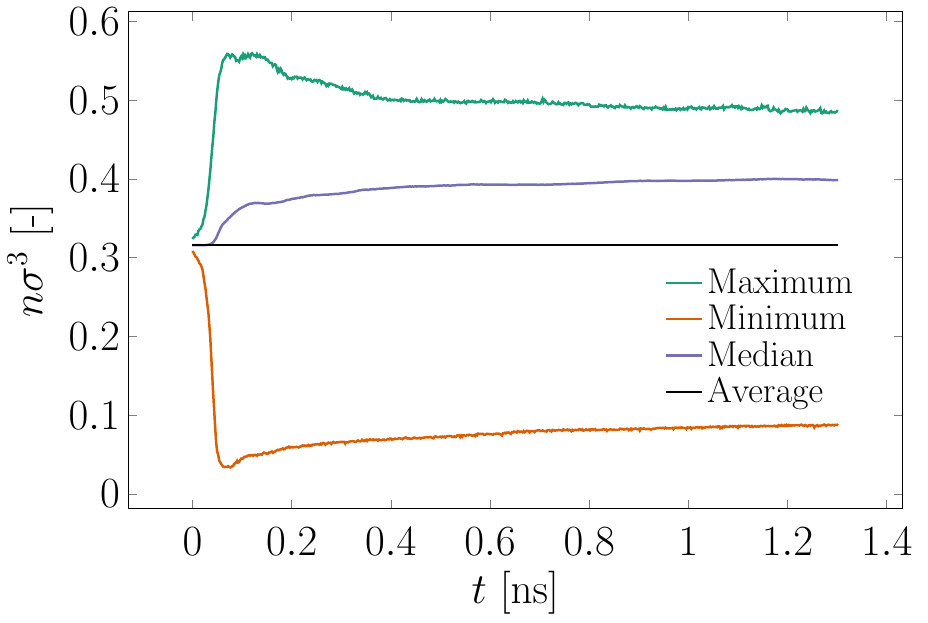}  }
  \caption{ }
  \end{subfigure}
   \hfill
  \begin{subfigure}[b]{0.44\columnwidth}
   \scalebox{0.37}{
  \includegraphics{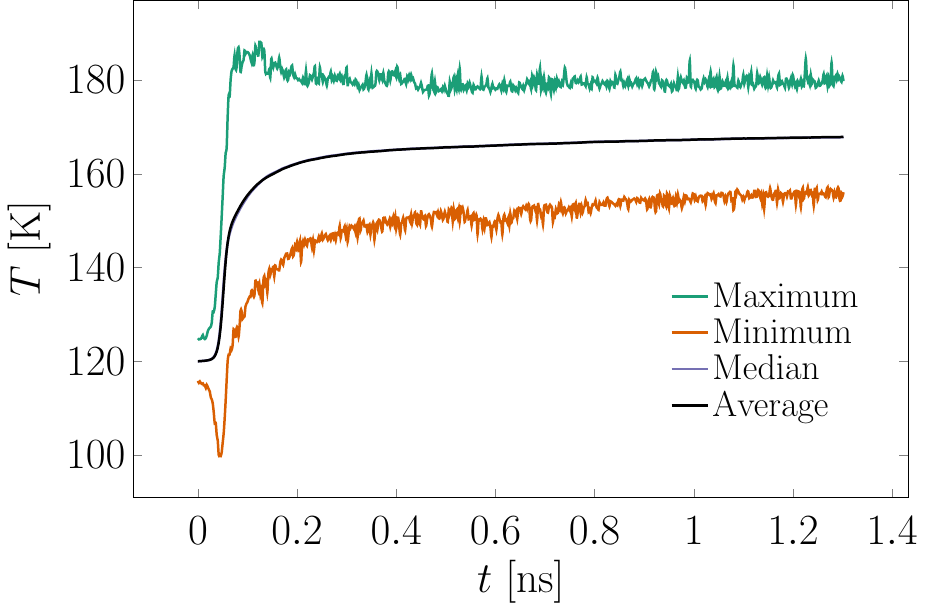}}
  \caption{ }
  \end{subfigure}
   \caption{Evolution of maximum, minimum, median and average values of normalized number density as well as temperature in the whole simulation domain $\Omega_x$ for spinodal decomposition of Argon with settings of case (C). Results are obtained using the DFP-SP model.}
   \label{fig21}
\end{figure}

\begin{figure}
  \centering
  \begin{subfigure}[b]{0.44\columnwidth}
   \scalebox{0.37}{
  \includegraphics{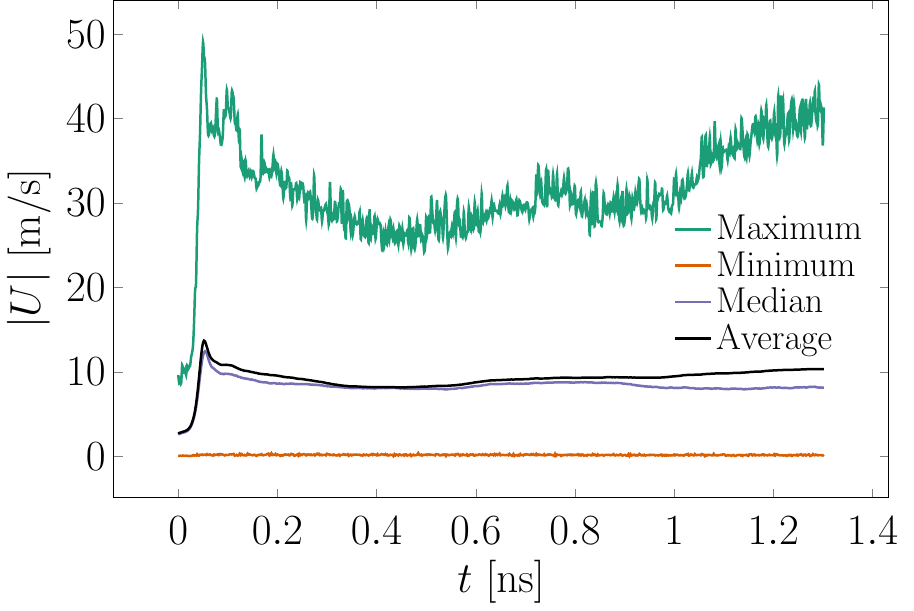}  }
  \caption{ }
    \label{fig:highdensvelo_high_den}
  \end{subfigure}
   \qquad
  \begin{subfigure}[b]{0.44\columnwidth}
   \scalebox{0.36}{
  \includegraphics{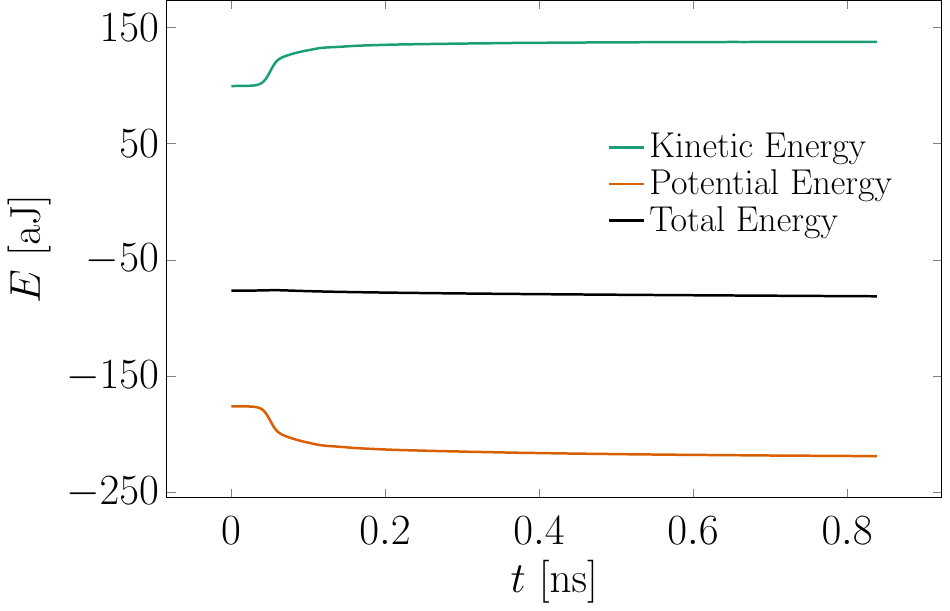}}
  \caption{ }
  \label{fig22b}
  \end{subfigure}
   \caption{
   Evolution of maximum, minimum, median and average values of velocity magnitude  in the whole domain $\Omega_x$ for the test case (C)  in simulating spinodal decomposition using the DFP-SP model are shown in (a). The evolution of kinetic, potential and total energy of the system are depicted in (b).
   }
   \label{fig22}
\end{figure}

\begin{figure}
  \centering
   \scalebox{0.85}{
  \includegraphics{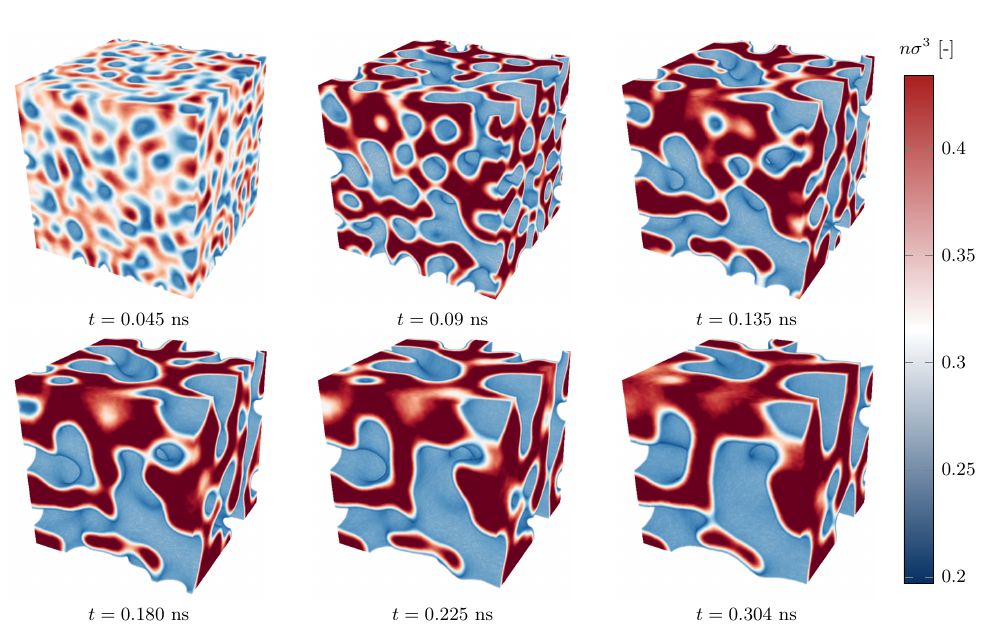}}
   \caption{
   Evolution of normalized number density for the spinodal decomposition of Argon in three-dimensional physical space $\Omega_x=[0,L]^3\ \mathrm{m}^3$ where $L=3\times10^{-8}\ \mathrm{m}$ obtained from simulation results of the DFP-SP model, with initial number density $n^{(0)}=8\times10^{27}\ \mathrm{m}^{-3}$ and temperature $T^{(0)}=\ 120\ \mathrm{K}$.
   }
   \label{fig23}
\end{figure}

\noindent Phase transition phenomena which are pronounced in breakup and coalescence of bubbles as well as droplets, comprise one of grand challenges in simulation of multi-phase flows. In particular, an accurate description of these processes demands incorporation of mass, momentum and energy transfer among different phases, if conventional multi-phase solvers   are considered \cite{shan1993lattice,watanabe2012phase}. However due to the fact that the considered EV kinetic framework provides a universal flow description across liquid and gas phases, a systematic exchange of mass, momentum, and energy at the interface is obtained naturally, consistent with the underlying molecular potential. In order to demonstrate the potential of the devised kinetic model in simulating such scenarios, here we study spinodal decomposition of Argon, where formation of droplets as well as bubbles can be investigated. 
\\ \ \\
\noindent Let us consider a simulation domain $\Omega_x=[0,L]^2\times[0,1]\ \mathrm{m}^3$ where $L=5\times10^{-8}\ \mathrm{m}$. Furthermore, assuming symmetry in the third dimension, we intend to study a two-dimensional (2D) planar fluid flow in $\Omega_x$ that, after a convergence study, is discretized with $250\times250\times1$ cells. As boundary conditions for relevant dimensions, i.e. $x_1$ and $x_2$, we consider specular reflection for particles that leave the domain. This leads to the Neumann boundary condition for the SP model owing to long-range interaction, i.e., $\tilde{n}_i \partial \tilde{\Phi}/\partial x_i  = 0$, where $\tilde{n}$ indicates the unit outward normal vector on the surface of the boundary $\partial \Omega_x$.
\\ \ \\
\noindent First, particles are initialized inside the domain $\Omega_x$ with a uniform distribution in physical space and Maxwellian in velocity for a given initial number density $n^{(0)}$ and temperature $T^{(0)}$. The idea is to pick an unstable state $(n^{(0)},T^{(0)})$ from a point on the phase diagram that does not belong to liquid nor vapour at the equilibrium. Then, we let the system relaxes towards the equilibrium. The expectation is that particles move away from this unstable initial condition until the corresponding mixture reaches a stable state at the equilibrium on the binodal diagram. Here we considered three test cases, as detailed in Table \ref{tab:spinodal}, with the main difference in their initial densities. The intention behind these cases is to study formation of droplets in (A) low density range and (B) medium density range. Furthermore formation of bubbles is tackled in a high density range test case of (C).
\begin{table}
  \begin{center}
\def~{\hphantom{0}}
  \begin{tabular}{lccccc}
      Test Case & $n^{(0)}\ [\mathrm{m^{-3}}]$ &$T^{(0)}\ [\mathrm{K}]$ & $\Delta t\ [\mathrm{s}]$ & $N_p$ \\[3pt]
       (A) \textit{Low density}    & $3\times 10^{27}$ & $120$ & $10^{-14}$ & $1.33 \times 10^{8}$\\
       (B) \textit{Medium density}  & $5\times 10^{27}$ & $120$ & $10^{-14}$ & $1.25 \times 10^{8}$\\
       (C) \textit{High density}    & $8\times 10^{27}$ & $120$ & $2\times 10^{-14}$ & $1.33 \times 10^{8}$
  \end{tabular}
  \caption{Simulation setting of the spinodal decomposition. Initial number density $n^{(0)}$, temperature $T^{(0)}$, the time step size $\Delta t$ and number of particles $N_p$ used in the simulations are reported here.}
  \label{tab:spinodal}
  \end{center}
\end{table}
\ \\ \ \\
    \textbf{Test case (A):}
 \textit{formation of nano droplets}\\
    As shown in Movie 1 and depicted in Fig.~\ref{fig13}, formation, and coalescence of droplets can be observed following the evolution of the number density. Note that two droplets, designated by rings around them and identified in the figure, are tracked in time for further analysis. 
    \\ \ \\
    As shown in Fig.~\ref{fig14}, unless interrupted by a merge with another droplet, the radius of both droplets first increases with the rate of $t^{1/3}$ and then their growth rate reduces to $t$. At the first stage, when the inertia of the fluid system is negligible, the Lifshitz-Slyozov mechanism is dominant \cite{siggia1979late}. This mechanism gives the growth rate of $t^{1/3}$ consistent with our result. Next, once we reach critical quenches (similar volume fraction of both phases), the growth rate stabilizes at the growth rate of $t$ \cite{siggia1979late,bastea1997spinodal}. 
    \\ \ \\
    Furthermore, statistics of number density, temperature, and velocity magnitude is shown in Figs.~\ref{fig15} and \ref{fig16a}. Moreover, we study the total energy of the system 
    \begin{equation}
    E_{\mathrm{total}}(t)=
    \underbrace{\frac{1}{2}\int_{\Omega_x} \rho(x,t) \mathbb{E}[V_j V_j|x,t] dx}_{E_{\mathrm{kin}}(t)}
    +
    \underbrace{\frac{1}{2m} \int_{\Omega_x} 
    \rho(x,t) \tilde{\Phi}(x,t) dx
   }_{E_{\mathrm{pot}}(t)}
\end{equation}
as derived in Appendix \S \ref{sec:Etot}, and here is evaluated numerically from the simulation results. As shown in Fig.~\ref{fig16b}, we observe energy transformation from potential to kinetic at the early stage of the spinodal decomposition, while the total energy stays almost constant. Furthermore, we observe relatively small decay of the total energy of the system, which is due to numerical errors in the deployed time integration scheme.
\ \\ \ \\
    \textbf{Test case (B):}
     \textit{formation of larger droplets}\\ Since here the initial unstable state has a larger density with respect to the case (A),  more sizable droplets are formed as illustrated in the evolution of the number density in Movie 2 and Fig.~\ref{fig17}. The maximum, median, and minimum values of number density, temperature, velocity magnitude as well as kinetic, potential, and total energy are computed. The results are shown in Figs.~\ref{fig18} and \ref{fig19b}. Unlike the test case (A), we observe the appearance of more nano droplets in the early stage of the spinodal decomposition, which leads to the subsequent formation of larger droplets as coalescence among more nano droplets occurs.
    \ \\ \ \\
    \textbf{Test case (C):}
     \textit{formation of nano bubbles}\\ Having initialized spinodal decomposition with a large enough density, here we observe the formation of bubbles instead of droplets, as illustrated in the evolution of the number density in Movie 3 and Fig.~\ref{fig20}. Similar to the last two test cases, statistical investigation of number density, temperature, velocity magnitude as well as kinetic, potential, and total energy are shown in Figs.~\ref{fig21} and \ref{fig22b}.
    \ \\ \ \\
    \textbf{Computational cost}:
     \ \\
    As expected, the computational cost of DFP-SP simulations remains identical, as the deployed spatial discretization as well as particle numbers among considered spinodal decomposition test cases are similar. However, once the liquid is formed, the average execution time per $100$ iterations ranges from $360~\mathrm{s}$ to $540~\mathrm{s}$, which can be explained by the in-homogeneity of the load balance as particles cluster inside the liquid phase. Note that here the parallelization is based on a domain decomposition of the physical space. 
    For the simulations of the spinodal decomposition, we deployed $40$ cores of Intel(R) Xeon(R) Platinum $8160$ CPU @ $2.10$GHz in parallel. 
 Using  this computational resource, the total computation time, e.g.  for case (C), with end time of $t_\mathrm{final}\approx1.3\ \mathrm{ns}$ 
 was about $4$ days.
 \\ \ \\
 \noindent To further demonstrate the capability of the devised DFP-SP model to address realistic and challenging multi-phase flows, here we extend the study of the spinodal decomposition to the three dimensional (3D) physical space. The simulation domain is set to $\Omega_x=[0,L]^3\ \mathrm{m}^3$ where $L=3\times10^{-8}\ \mathrm{m}$. Here, $\Omega_x$ is discretized by $100^3$ cells. As boundary conditions, specular reflection for particles leaving the domain as well as the Neumann boundary condition for the SP model of long-range interaction, is considered.
Here, $N_p=4.32\times10^8$ particles and time step size of $\Delta t= 2\times 10^{-14}$~s are considered. Initially, we sample the position of particles inside $\Omega_x$ from a uniform distribution with a weight that provides us with the initial number density of $n^{(0)}=8\times10^{27}\ \mathrm{m^{-3}}$. Furthermore, similar to the 2D case, the velocity of particles are initialized by sampling from the Maxwellian distribution such that we obtain the initial temperature of $T^{(0)}=120\ \mathrm{K}$. 
\\ \ \\
Having employed similar computational resources like the one for 2D case, we observe that the average execution time for $50$ iterations per cell approximately ranges from $ 600~\mathrm{s}$ to $2040~\mathrm{s}$. Here, the effect of load imbalance due to particle clustering is more pronounced. Hence, we acknowledge the need for devising an efficient load balancing scheme that benefits from density distribution in the physical space, in parallel computations of multi-phase flows using particles. Similar to the test case (C), as shown by the evolution of number density in Fig.~\ref{fig23} and more profoundly in Movie 4, we observe the formation of bubbles contained in the liquid.  Simulation result of this spinodal decomposition using DFP-SP model  was carried out until $t_{\mathrm{final}}\approx0.304\ \mathrm{ns}$ with time step size of $\Delta t=0.02\ \mathrm{ps}$ which took about $3$ days.
{\ \\  \ \\\color{ms3}
Finally, we provide a rough estimate on the computational cost of the devised DFP-SP in comparison to standard MD algorithm, for a fixed statistical noise and time step size. Given the volume of the domain and the number density, there are $N_\text{mol}=216\times 10^{3}$ physical molecules inside the cube. In order to control the noise in observables, e.g. bulk velocity, the MD simulation should be averaged over an ensemble of $N_{\mathrm{ens}}$ snapshots.  By repeating the MD simulation $N_\text{ens}$ times, one can obtain the  variance of $\epsilon^2\propto 1/(N_\text{mol}N_\text{ens})$. In the case of our particle solver, we deployed $N_p=4.32\times 10^{8}$ samples which lead to the noise of $\epsilon^2\propto 1/N_p=2.3\times 10^{-9}$. To obtain similar noise with MD, one would need to use  $N_\text{ens}\approx 2000$ samples. Hence the total number of operations for one time step of MD becomes $N_\text{ens}N_\text{mol}^2\approx 9\times 10^{13}$.
      However, the number  of operations for DFP-SP is linear with $  N_p =4.32\times 10^{8}$ resulting in a significant speed-up compared to MD. {Note that while introducing a cut-off leads to faster MD simulations, still the cost remains significantly higher than DFP-SP as the latter does not explicitly compute short-range interactions.}
}


 \section{Conclusion and outlook}
\label{sec:conclusion}
\noindent Multi-phase flows far from the equilibrium demand for high-fidelity models beyond the conventional hydrodynamics. This study lays out a Fokker-Planck-Poisson approach which systematically reduces the multi-phase kinetic description into a set of SDEs suitable for efficient particle simulations. The short-range molecular collisions are described by the dense FP model, whereas the long-range interactions are cast into a SP elliptic model. The resulting system then is simulated using a particle Monte-Carlo scheme in combination with a fast elliptic solver. First the accuracy and performance of the devised DFP-SP model are assessed in predicting the surface tension and the evaporation rate of a liquid slab once in contact with the vacuum and once in contact with the equilibrium vapor phase. Furthermore, two intriguing multi-phase problems were addressed. The inverted-temperature gradient phenomenon which is observed in the vapor phase contained between two liquid slabs kept at slightly different temperatures, was accurately reproduced by the DFP-SP model. Further insights into the phenomenon were gained by analyzing single droplet evaporation process. Next, the spinodal decomposition of Argon was investigated at different initial densities. Formation and coalescence of bubbles as well as droplets are illustrated. Moreover, the droplet growth rates are found to be consistent with the Lifschitz-Slyozov solution. Future steps will focus on integration of more accurate molecular potentials such as the Lennard-Jones potential into the Fokker-Planck-Poisson framework. We expect that this can be achieved by employing a system of SP equations. Moreover, extension of the model towards polyatomic degrees-of-freedom, mixtures and chemical reactions will be pursued.      

\section*{Acknowledgements}
\noindent 
The authors would like to thank the anonymous reviewers for their valuable and constructive comments and suggestions which improved the quality of the paper. The authors gratefully acknowledge the supports of Manuel Torrilhon throughout this study. Furthermore, we appreciate the stimulating discussions with Patrick Jenny.

\appendix
\section{ESMC-Vlasov solution algorithm}
\label{app:ESMC_Vlasov_alg}
\ \\
Here first the Monte-Carlo approach to account for the Enskog equation i.e. ESMC is reviewed. Then, the direct approach in solving the Vlasov integral is explained. At the end, the combined ESMC-Vlasov algorithm is presented which is deployed in this work as the benchmark solution of the EV equation. For more details on DSMC approach for flow simulation based on the EV equation see \cite{frezzotti2019direct}.
\\ \ \\
Consider the position $X$ and the velocity $V$ as random variables which describe the evolution of $\mathcal{F}$ in the EV equation. As a Monte-Carlo particle method, instead of solving for $\mathcal{F}$ directly, the samples of  position and velocity are evolved. Hence, at each time step first the velocity of particles are evolved following a jump process. Then, particles with new velocities are streamed freely till the end of time step.
\\
\subsection{ESMC collision step}
\ \\
The modified ESMC solution algorithm was introduced to mimic the solution algorithm  of classical Bird's DSMC method with the modification for collision frequency as well as incorporating the fact that the colliding pair might lie in different cells.
 Using Bird's No-Time-Counter (NTC) approach \cite{Bird}, particles collide with the probability of
\begin{flalign}
\mathcal{P}_{\text{coll.}} = \frac{\sigma_T c_r Y(n_{\mathrm{mid}})}{\{\sigma_T c_r   Y(n_{\mathrm{mid}})\}_\text{max}} 
\end{flalign}
where $\sigma_T=\pi \sigma^2$ is the hard-sphere collision cross-section, $g_r$ the magnitude of relative velocity for a colliding pair with number density of $n_{\mathrm{mid}}$  at their midpoint. Here, the subscript max indicates an upper limit for the values of $\sigma_T c_r$  in the the computational cell.  Particle pair with the index $(i,j)$ collides while satisfying the conservation laws. Bringing the randomness in the post-collision direction of velocity, the velocity of the colliding pair during a collision evolves as
\begin{flalign}
{V}^{(i)} &=   V_{\text{c}} + \frac{{g}}{2} \ \ \text{and} \label{eq:dsmc_coll_1}\\
{V}^{(j)} &=   V_{\text{c}} - \frac{{g}}{2}, \ \ \text{where}\label{eq:dsmc_coll_2}\\
{g} &=  g_r \left[\cos(\theta), \sin(\theta)\cos(\phi), \sin(\theta)\sin(\phi)\right]^T. \label{eq:dsmc_coll_3}
\end{flalign}
Consistent with the hard-sphere model, the angles of the post-collision relative velocity $g$ are drawn randomly with the uniform probability on the surface of  the ball $\mathcal{B}(0,1)$ which is centered at zero and has the unit radius. Due to efficiency in sampling from one dimensional uniform distribution in practice, the angles $\theta$ and $\phi$ can be easily sampled using the map
\begin{flalign}
\theta &= \mathrm{arccos}( 2 \alpha_1-1 ) \label{eq:dsmc_coll_4} \ \  \text{and}\\
\phi &= 2 \pi \alpha_2~,
\label{eq:dsmc_coll_5}
\end{flalign}
where $\alpha_{1,2} \sim \mathcal{UR}([0,1]) $, i.e. continuous uniform distribution with support in the interval $[0,1]$. The velocity of center of mass  $V_{\text{c}}:= ( V^{(i)}+ V^{(j)})/2$ and the magnitude of relative velocity $ g:=  V^{(i)}- V^{(j)}$, i.e. $g_r:=|g|$, remain constant during elastic collisions. Let  $\hat{k}$ indicate the unit vector with the direction pointing to the center of  $j$th particle from $i$th particle's center of mass. Hence, the number of collision candidates $N_{\mathrm{cand}}^{(I)}$ for the time step size $\Delta t$ in the cell with index $I$ and $N_p^{(I)}$ particles is
\begin{flalign}
N_{\mathrm{cand}}^{(I)} &= \frac{1}{2} \omega_{\mathrm{max}}^{(I)} N_p^{(I)},\ \ \textrm{where}\\
w^{(I)}_{\text{max}} &= 4 \pi \sigma^2 \max_{(ij)}\{(g,k) n^{(J)} Y (n_{\textrm{mid}})  \} \Delta t~.
\end{flalign}
We initialize the upper bound of the collision rate using the thermal velocity
\begin{flalign}
\omega_{\textrm{max},0}^{(I)} &= 4 \pi \sigma^2 n^{(I)} \Theta_0^{(I)}  Y(n^{(I)})  \Delta t ,\ \ \textrm{where}\\
\Theta_0^{(I)} &= 10 \sqrt{\frac{k T_0^{(I)}}{m}}~,
\end{flalign}
and update it every time step if a collision with $\omega^{(ij)}>w^{(I)}_{\text{max}}$ where
\begin{flalign}
\omega^{(ij)}&:=4 \pi \sigma^2 (g,\hat{k}) n^{(J)} Y (n_{\mathrm{mid}}) \Delta  t \nonumber
\end{flalign}
occurs. Details of the implemented solution algorithm is described in Alg.~\ref{alg:ESMC}. Note that here, $\mathcal{UI}([a,b])$ for $a,b \in \mathbb{N}$ and $b\geq a$ indicates a uniform discrete distribution over the set of natural numbers between $a$ and $b$. Furthermore, simply looping over cells and performing Enskog collisions similar to the implementation of DSMC may lead to introducing bias  due to outside cell collisions and the fact that particles of one cell always  are considered for collision before another. Therefore, an unbiased particle selection algorithm is devised here for the ESMC algorithm.
\begin{algorithm}
\caption{Collision step of ESMC}
\label{alg:ESMC}
\begin{algorithmic}
\State{Compute total number of collision candidates $N_{\mathrm{cand}}^{\mathrm{tot}}=\sum_{I=1}^{N_{\mathrm{cells}}} N_{\mathrm{cand}}^{(I)}$. }
\State{Create discrete accumulative distribution set $f = \{ N_{\mathrm{cand}}^{(1)},N_{\mathrm{cand}}^{(1)}+N_{\mathrm{cand}}^{(2)},..., N_{\mathrm{cand}}^{\mathrm{tot}} \}$}.
\For{$l=1,...,N_{\mathrm{cand}}^{\mathrm{tot}}$}
\State{Draw $m\sim \mathcal{UI}([1,N_\mathrm{cand}^{\mathrm{tot}}])$.}
\State{Let $I$ indicate the smallest index for which $m\leq f_I$.}
\State{Overwrite $f$ such that $f_q=f_q-1$ for $q=I,...,N_\mathrm{cells}$.}
\State{Pick random particle $i\sim \mathcal{UI}([1,N_{p}^{(I)}])$.}
\State{Draw a random direction $\hat{k}\sim \mathcal{UR}(\partial \mathcal{B}(x^{(i)},1))$.}
\State{Let $J$ denote the index of cell that contains the point $x^{(i)}+\sigma \hat{k}$.}
\State{Draw particle $j\sim \mathcal{UI}([1,N_{p}^{(J)}])$.}
\If{$\omega^{(ij)}> \omega_{\mathrm{max}}^{(I)}$}
\State{Update $\omega_{\mathrm{max}}^{(I)}=\omega^{(ij)}$.}
\EndIf
\State{Draw $r\sim \mathcal{UR}([0,1])$.}
\If{$\omega^{(ij)}/\omega^{(I)}_\mathrm{max}<r$}
\State{Collide particle $i$ with $j$ using Eqs.~\eqref{eq:dsmc_coll_1}-\eqref{eq:dsmc_coll_2}.}
\EndIf
\EndFor
\\
\Return
\end{algorithmic}
\end{algorithm}

\subsection{Numerics of long-range interaction}
\label{sec:num_long_range}
\ \\
Similar to the SP model, first the Vlasov integral is estimated numerically on the faces of the computational cell indexed $(i,j,k)$. It is simpler to take the integral in spherical coordinates and read the density of the quadrature points from the density field. Then, the numerical estimate of the Vlasov integral with origin at the center of the face that is indexed by $f_l$ becomes  
\begin{flalign}
\label{eq:integ-sp1}
U_{f^{(i,j,k)}_l}^{(n)}     &=  \phi_0 \sum_{\alpha=1}^{N_r} \sum_{\beta=1}^{N_\mathrm{\phi}}  
\sum_{\gamma=1}^{N_\mathrm{\theta}}
\left(\frac{\sigma}{r^{(\alpha)}}\right)^6 n (y^{(\alpha,\beta,\gamma)}) (r^{(\alpha)})^2 \sin(\phi^{(\beta)})  \Delta r^{(\alpha)}  \Delta \phi^{(\beta)} \Delta \theta^{(\gamma)},\\
y^{(\alpha,\beta,\gamma)} &= \Big( r^{(\alpha)} \cos(\theta^{(\gamma)}) \sin(\phi^{(\beta)}), r^{(\alpha)} \sin(\theta^{(\gamma)}) \sin(\phi^{(\beta)}), r^{(\alpha)} \cos(\phi^{(\beta)}) \Big)^T.
\end{flalign}
Let us combine the variables in a vector
$x=[r, \phi, \theta]^T$ with lower and upper bounds in vectors $a=[\sigma, 0, 0]^T$ and $b=[r_{\mathrm{cut}}, \pi, 2\pi]^T$, respectively.  For numerical integration, Legendre-Gauss quadrature rule using $N_r=N_\phi=N_\theta=10$ are employed here. The $\eta$'th quadrature weight $w$ and point $\xi$ from the standard interval $[-1,1]$ are simply mapped back to the interval of interest for the $s$'th component of $x$ via
\begin{numcases}{}
x_{s}^{(\eta)} = \xi^{(\eta)} \frac{b_s-a_s}{2} +\frac{a_s+b_s}{2}\ \ \ \textrm{and}\\
\Delta x_{s}^{(\eta)} = w^{(\eta)} \frac{b_s-a_s}{2}.
\end{numcases}
Once the potential on the faces are computed numerically, the attractive forces are estimated via finite differences.
\subsection{ESMC-Vlasov algorithm}
To combine long-range interaction along with the short-range one, first at each time step, the ESMC collision step is carried out. Then, the  velocities are updated to include attractive contribution once the numerical integration of the Vlasov on the faces are evaluated. Algorithm~\ref{alg:esmc_vlasov} shows the outline of the the ESMC-Vlasov solution algorithm used as the benchmark in this study.
\begin{algorithm}
\caption{ESMC-Vlasov algorithm}
\label{alg:esmc_vlasov}
\begin{algorithmic}
\While{$t<T$}
\State{Compute post-collision velocity $V^{(*)}$ following ESMC collision Algorithm~\ref{alg:ESMC}.}
\State{Compute Vlasov integral numerically on  faces of each cell by Eq.~\eqref{eq:integ-sp}.}
\State{Update the estimated velocity by including attractive forces by Eq.~\eqref{eq:up_vel_attraction}.}
\State{Stream position of all particles $X^{(n+1)}=X^{(n)}+V^{(n+1)}\Delta t$.}
\State {Apply the boundary conditions.}
\State{Increment $t$ by $\Delta t$.}
\EndWhile
\\
\Return
\end{algorithmic}
\end{algorithm}

\section{Hybridized Discontinuous Galerkin Solver}
\label{sec:HDGS}
\noindent The solver used for the screened Poisson equation in the parallel high performance computing (HPC) case is based on the hybridized Discontinuous Galerkin solver in the particle code PICLas~\cite{fasoulas2019}. This solver uses a spectral element representation and is therefore referred to as hybridized Discontinuous Galerkin spectral element method (HDGSEM). The main advantage of this method is its reasonably high performance in parallelization. Here, an overview and introduction of the concept is given. The detailed derivation of the method as well as its application in simulations of plasma physics can be found in \cite{pfeiffer2019particle}.
\\ \ \\
As a first step, the screened Poisson equation \eqref{eq:sp_bounded} is decomposed into a first order system of partial differential equations, i.e.,
\begin{equation}
\lambda^2 u -  \frac{\partial q_i}{\partial x_i}  = n,\quad
  q_i  = \frac{\partial u}{\partial x_i},
  \label{eq:poisson_decomposed}
\end{equation}
with Dirichlet and Neumann boundary conditions encoded by
\begin{equation}
\left.u\right|_{\partial \Omega_\mathrm{Dir}}=u_\mathrm{Dir},\quad 
\left.  q_i  \tilde{n}_i\right|_{\partial \Omega_\mathrm{Neu}}={( q_i  \tilde{n}_i)}_\mathrm{Neu}.
\end{equation}
Here, $\tilde{n}$ denotes outward unit normal vector on the boundary surface of the domain. One can obtain the variational form by multiplying the equations with test functions $\overline{u}$ and $\overline{  q}$, respectively, and integrate over the volume of each element $K$, i.e., 
\begin{flalign}
 \langle \lambda^2 u,\overline{u}\ \rangle_K -  \langle \frac{\partial q_i}{\partial x_i},\overline{u} \rangle_K  &=  \langle n,\overline{u} \rangle_K  \\
 \langle  q_i,\overline{  q}_i\ \rangle_K  &=  \langle \frac{\partial u}{\partial x_i} ,\overline{  q}_i \rangle_K. 
\end{flalign}
{Here, $\langle a_i, b_i\rangle_K$ denotes the integration of the inner-product $a_ib_i$ inside the K'th element.}
Next, these equations are integrated by parts and discontinuous solutions at element interfaces are allowed for $  q$ by applying jump condition. The main idea of the HDG method is to introduce an additional unknown $\theta$, which is unique on each element interface, in order to reduce dimension of the global system  to the number of interfaces.
By skipping several steps in the derivation, the final system of equations including the auxiliary variable $\theta$ as well as stabilization parameter $\tau>0$,   see \cite{cockburn2009hybridizable}, can be obtained as 
\begin{flalign}
\langle \lambda^2 u,\overline{u}\rangle _K + \langle   \frac{\partial q_i}{\partial x_i},  \overline{u}\rangle _K 
-\sum_{e\in\partial K}\{\tau u \overline{u}\}_e +\sum_{e\in\partial K}\{\tau \theta \overline{u}\}_e  &= \langle n,\overline{u}\rangle _K,
\label{eq:hdgsystem1}\\
\langle  q_i,\overline{ q}_i\rangle _K  + \langle  u, \frac{\partial \overline{ q}_i}{\partial x_i}\rangle _K
-\sum_{e\in\partial K}\{\theta \overline{ q}_i \tilde{n}_i\}_e &=0, 
\label{eq:hdgsystem2}\\
\textrm{and}\ \ \ 
\sum_{e\in\partial K, K'}\left(\{ q_i \tilde{n}_i \overline{\theta}\}_e +
\{\tau u \overline{\theta}\}_e - \{\tau \theta \overline{\theta}\}_e \right) &= 
\sum_{e\in\Omega_\mathrm{Neu}} \{ q_i  \tilde{n}_i \overline{\theta}\}_e,
\label{eq:hdgsystem3}
\end{flalign}
with  the adjacent element $K'$ on the interface indicated by $e$. {Note that here $\{s\}_e$ indicates \textit{average} of $s$ on the edge $e$, see \cite{arnold2002unified}}. For the sake of simplicity, the transformation to/from the reference space is not explained here. This transformation as well as a detailed derivation of Eqs.~\eqref{eq:hdgsystem1}-\eqref{eq:hdgsystem3} can be found in \cite{pfeiffer2019particle}.
\\ \ \\
In order to find the solution $u$ of Eq.~\eqref{eq:poisson_decomposed}, first, a global system of equations 
is solved for the auxiliary variable $\theta$ only on the surfaces using a conjugate-gradient method. The final volume solution $u$ inside each cell is calculated in a post processing step locally in each element. Once the solution $u$ with expansion in polynomial basis functions inside each cell is found, its derivatives needed for computing attractive force  $H$ at any point inside the cell can be obtained simply by applying derivatives on the basis functions.
\section{Total Energy for Enskog-Vlasov System}
\label{sec:Etot}
\noindent In order to derive evolution of the total energy in the EV equation, let us first multiply Eq.~\eqref{eq:full_kinetic_model} by $ v_jv_j/2$ and integrate over $v$. Therefore we get
\begin{eqnarray}
\frac{\partial}{\partial t}\int_{\mathbb{R}^3} \frac{1}{2}\mathcal{F} v_jv_jdv+\frac{\partial }{\partial x_i}\left(\frac{1}{2}\int_{\mathbb{R}^3} v_i v_jv_j\mathcal{F} dv\right)-\int_{\mathbb{R}^3} v_j\frac{H_j}{m}\mathcal{F}dv&=&0.
\end{eqnarray}
Using the notation
\begin{eqnarray}
 \mathbb{E}[Q(V|x,t)]=\frac{1}{\rho}\int_{\mathbb{R}^3} Q(v)\mathcal{F}(v,x,t)dv,
\end{eqnarray}
yields
\begin{eqnarray}
\frac{\partial}{\partial t}\left(\frac{1}{2}\rho \mathbb{E}[V_jV_j|x,t]\right)+\frac{\partial}{\partial x_i}\left(\frac{1}{2}\rho \mathbb{E}[V_iV_jV_j|x,t]\right)-\frac{H_j}{m}\rho \mathbb{E}[V_j|x,t]&=&0.
\end{eqnarray}
Now by taking integral over the physical space $\Omega$ we get
\begin{eqnarray}
\frac{\partial}{\partial t}\int_\Omega \left(\frac{1}{2}\rho \mathbb{E}[V_jV_j|x,t]dx\right)+\int_{\partial \Omega}\left(\frac{1}{2}\rho \mathbb{E}[V_iV_jV_j|x,t]n_ids\right)&&\nonumber\\ +\int_{\partial \Omega}\left(\frac{\Phi}{m}\rho \mathbb{E}[V_j|x,t]n_jds\right)+\int_\Omega \frac{\Phi}{m}\frac{\partial \rho}{\partial t}dx &=&0,
\end{eqnarray}
where $\partial \Omega$ is the boundary of $\Omega$ and $ n$ its normal. Therefore for the interval $t=0$ to $t=t_f$ we get
\begin{eqnarray}
E_\mathrm{kin}(t_f)&=&E_\mathrm{kin}(0)-\int_0^{t_f}\int_\Omega \frac{\Phi}{m}\frac{\partial \rho}{\partial t}dxdt \nonumber \\
&&-\int_0^{t_f} \int_{\partial \Omega} \left(\frac{1}{2}\mathbb{E}[V_iV_jV_j|x,t]+\frac{\Phi}{m}\mathbb{E}[V_i|x,t]\right)\rho n_idsdt,
\end{eqnarray}
where $E_\mathrm{kin}=\int_\Omega 1/2\rho\mathbb{E}[V_jV_j|x,t] dx$. The surface term can be computed by averaging over particles crossing the boundary of the domain. If we have the homogeneous boundary condition however, it simplified to 
\begin{eqnarray}
\label{eq:homo}
E_\mathrm{kin}(t_f)=E_\mathrm{kin}(0)-\int_0^{t_f}\int_\Omega \frac{\Phi}{m}\frac{\partial \rho}{\partial t}dxdt.
\end{eqnarray}
Since $\Phi$ depends on $\rho$ through a time independent symmetric kernel 
\begin{eqnarray}
\Phi( x,t)&=&\int_\Omega \mathcal{K}( x^\prime, x) \rho( x^\prime,t) dx^\prime,
\end{eqnarray}
we obtain 
\begin{eqnarray}
\frac{\partial \Phi}{\partial t}&=&\int_\Omega \mathcal{K} \frac{\partial \rho}{\partial t} dx
\end{eqnarray}
and thus
\begin{eqnarray}
\int_\Omega \rho \frac{\partial \Phi}{\partial t} dx&=&\int_\Omega \rho( x) \int_\Omega \mathcal{K}( x^\prime , x) \frac{\partial \rho ( x^\prime)}{\partial t} dx^\prime dx \nonumber \\&=&\int_\Omega \frac{\partial \rho( x^\prime)}{\partial t} \int_\Omega \mathcal{K}( x^\prime  , x) \rho( x) dx dx^\prime \\
&=&\int_\Omega \frac{\partial \rho( x^\prime)}{\partial t} \int_\Omega \mathcal{K}( x  , x^\prime) \rho( x) dxdx^\prime \\
&=&\int_\Omega \frac{\partial \rho }{\partial t}\Phi dx^\prime.
\end{eqnarray}
Therefore Eq.~\eqref{eq:homo} results in the following energy conservation  
\begin{eqnarray}
E_\mathrm{kin}(t_f)+E_\mathrm{pot}(t_f)=E_\mathrm{kin}(0)+E_\mathrm{pot}(0),
\end{eqnarray}
where $E_\mathrm{pot}=1/2\int_\Omega \rho \Phi/m dx$ is the potential energy. 
\bibliographystyle{elsarticle-num}
\bibliography{refereneces}

\begin{thebibliography}{10}
\expandafter\ifx\csname url\endcsname\relax
  \def\url#1{\texttt{#1}}\fi
\expandafter\ifx\csname urlprefix\endcsname\relax\def\urlprefix{URL }\fi
\expandafter\ifx\csname href\endcsname\relax
  \def\href#1#2{#2} \def\path#1{#1}\fi

\bibitem{worner2012numerical}
M.~W{\"o}rner, Numerical modeling of multiphase flows in microfluidics and
  micro process engineering: a review of methods and applications,
  Microfluidics and nanofluidics 12 (2012) 841--886.

\bibitem{schwarzkopf2011multiphase}
J.~D. Schwarzkopf, M.~Sommerfeld, C.~T. Crowe, Y.~Tsuji, Multiphase flows with
  droplets and particles, CRC press, 2011.

\bibitem{benson2004kinetic}
C.~M. Benson, D.~A. Levin, J.~Zhong, S.~F. Gimelshein, A.~Montaser, Kinetic
  model for simulation of aerosol droplets in high-temperature environments,
  Journal of thermophysics and heat transfer 18~(1) (2004) 122--134.

\bibitem{alharthy2013multiphase}
N.~S. Alharthy, T.~Nguyen, T.~Teklu, H.~Kazemi, R.~Graves, et~al., Multiphase
  compositional modeling in small-scale pores of unconventional shale
  reservoirs, in: SPE Annual Technical Conference and Exhibition, Society of
  Petroleum Engineers, 2013, p. D031S052R008.

\bibitem{hudgins2016neutral}
D.~Hudgins, N.~Gambino, B.~Rollinger, R.~Abhari, Neutral cluster debris
  dynamics in droplet-based laser-produced plasma sources, Journal of Physics
  D: Applied Physics 49~(18) (2016) 185205.

\bibitem{Bird}
G.~A. Bird, Molecular gas dynamics and the direct simulation of gas flows,
  Clarendon Press, 1994.

\bibitem{Chapman1953}
S.~Chapman, T.~G. Cowling, The mathematical theory of non-uniform gases: an
  account of the kinetic theory of viscosity, thermal conduction and diffusion
  in gases, Cambridge university press, 1970.

\bibitem{Cercignani}
C.~Cercignani, The {Boltzmann} Equation and Its Applications, Springer, 1988.

\bibitem{enskog1922kinetische}
D.~Enskog, Kinetische Theorie der W{\"a}rmeleitung: Reibung und
  Selbst-diffusion in Gewissen verdichteten gasen und fl{\"u}ssigkeiten,
  Almqvist \& Wiksells boktryckeri-a.-b., 1922.

\bibitem{van1973modified}
H.~Van~Beijeren, M.~H. Ernst, The modified enskog equation, Physica 68 (1973)
  437--456.

\bibitem{resibois1978h}
P.~Resibois, H-theorem for the (modified) nonlinear enskog equation, Journal of
  Statistical Physics 19 (1978) 593--609.

\bibitem{vlasov1978many}
A.~Vlasov, Many-particle theory and its application to plasma, New York, Gordon
  and Breach, 1978.

\bibitem{karkheck1981kinetic}
J.~Karkheck, G.~Stell, Kinetic mean-field theories, The Journal of Chemical
  Physics 75~(3) (1981) 1475--1487.

\bibitem{grmela1971kinetic}
M.~Grmela, Kinetic equation approach to phase transitions, Journal of
  Statistical Physics 3~(3) (1971) 347--364.

\bibitem{landau1959fluid}
L.~D. Landau, E.~M. Lifshitz, Fluid mechanics, flme.

\bibitem{espanol2003smoothed}
P.~Espanol, M.~Revenga, Smoothed dissipative particle dynamics, Physical Review
  E 67~(2) (2003) 026705.

\bibitem{liu2007dissipative}
M.~Liu, P.~Meakin, H.~Huang, Dissipative particle dynamics simulation of
  multiphase fluid flow in microchannels and microchannel networks, Physics of
  Fluids 19~(3) (2007) 033302.

\bibitem{broadwell_1964}
J.~E. Broadwell, Study of rarefied shear flow by the discrete velocity method,
  Journal of Fluid Mechanics 19~(3) (1964) 401--414.

\bibitem{gamba2009spectral}
I.~M. Gamba, S.~H. Tharkabhushanam, Spectral-lagrangian methods for collisional
  models of non-equilibrium statistical states, Journal of Computational
  Physics 228~(6) (2009) 2012--2036.

\bibitem{wu2013deterministic}
L.~Wu, C.~White, T.~J. Scanlon, J.~M. Reese, Y.~Zhang, Deterministic numerical
  solutions of the {Boltzmann} equation using the fast spectral method, Journal
  of Computational Physics 250 (2013) 27--52.

\bibitem{wu2015fast}
L.~Wu, Y.~Zhang, J.~M. Reese, Fast spectral solution of the generalized enskog
  equation for dense gases, Journal of Computational Physics 303 (2015) 66--79.

\bibitem{wu2016non}
L.~Wu, H.~Liu, J.~M. Reese, Y.~Zhang, Non-equilibrium dynamics of dense gas
  under tight confinement, Journal of Fluid Mechanics 794 (2016) 252--266.

\bibitem{kremer1988enskog}
G.~Kremer, E.~Rosa~Jr, On {Enskog}'s dense gas theory. i. the method of moments
  for monatomic gases, The Journal of chemical physics 89~(5) (1988)
  3240--3247.

\bibitem{torrilhon2016modeling}
M.~Torrilhon, Modeling nonequilibrium gas flow based on moment equations,
  Annual review of fluid mechanics 48 (2016) 429--458.

\bibitem{struchtrup2019grad}
H.~Struchtrup, A.~Frezzotti, Grad’s 13 moments approximation for
  {Enskog}-{Vlasov} equation, in: AIP Conference Proceedings, Vol. 2132, AIP
  Publishing LLC, 2019, p. 120007.

\bibitem{wang2020kinetic}
P.~Wang, L.~Wu, M.~T. Ho, J.~Li, Z.-H. Li, Y.~Zhang, The kinetic
  shakhov--enskog model for non-equilibrium flow of dense gases, Journal of
  Fluid Mechanics 883.

\bibitem{heylmun2019quadrature}
J.~C. Heylmun, B.~Kong, A.~Passalacqua, R.~O. Fox, A quadrature-based moment
  method for polydisperse bubbly flows, Computer Physics Communications 244
  (2019) 187--204.

\bibitem{Bird1963}
G.~A. Bird, Approach to translational equilibrium in a rigid sphere gas,
  Physics of Fluids 6~(10) (1963) 1518--1519.

\bibitem{montanero1997simulation}
J.~M. Montanero, A.~Santos, Simulation of the {Enskog} equation a la {Bird},
  Physics of Fluids 9~(7) (1997) 2057--2060.

\bibitem{Alexander1995}
F.~J. Alexander, A.~L. Garcia, B.~J. Alder, A consistent {Boltzmann} algorithm,
  Physical Review Letters 74~(26) (1995) 5212.

\bibitem{hadjiconstantinou2000surface}
N.~G. Hadjiconstantinou, A.~L. Garcia, B.~J. Alder, The surface properties of a
  {Van der Waals} fluid, Physica A: Statistical Mechanics and its Applications
  281~(1-4) (2000) 337--347.

\bibitem{Frezzotti1997}
A.~Frezzotti, A particle scheme for the numerical solution of the {Enskog}
  equation, Physics of Fluids 9~(5) (1997) 1329--1335.

\bibitem{frezzotti2019direct}
A.~Frezzotti, P.~Barbante, L.~Gibelli, Direct simulation monte carlo
  applications to liquid-vapor flows, Physics of Fluids 31~(6) (2019) 062103.

\bibitem{Jenny2010}
P.~Jenny, M.~Torrilhon, S.~Heinz, A solution algorithm for the fluid dynamic
  equations based on a stochastic model for molecular motion, Journal of
  Computational Physics 229~(4) (2010) 1077--1098.

\bibitem{sadr2017continuous}
M.~Sadr, M.~H. Gorji, A continuous stochastic model for non-equilibrium dense
  gases, Physics of Fluids 29~(12) (2017) 122007.

\bibitem{frezzotti2005mean}
A.~Frezzotti, L.~Gibelli, S.~Lorenzani, Mean field kinetic theory description
  of evaporation of a fluid into vacuum, Physics of Fluids 17~(1) (2005)
  012102.

\bibitem{piechor1994discrete}
K.~Piech{\'o}r, Discrete velocity models of the {Enskog}-{Vlasov} equation,
  Transport Theory and Statistical Physics 23 (1994) 39--74.

\bibitem{frezzotti2018mean}
A.~Frezzotti, L.~Gibelli, D.~Lockerby, J.~Sprittles, Mean-field kinetic theory
  approach to evaporation of a binary liquid into vacuum, Physical Review
  Fluids 3 (2018) 054001.

\bibitem{sadr2019treatment}
M.~Sadr, M.~H. Gorji, Treatment of long-range interactions arising in the
  {Enskog}--{Vlasov} description of dense fluids, Journal of Computational
  Physics 378 (2019) 129--142.

\bibitem{he2002thermodynamic}
X.~He, G.~D. Doolen, Thermodynamic foundations of kinetic theory and lattice
  {Boltzmann} models for multiphase flows, Journal of Statistical Physics
  107~(1-2) (2002) 309--328.

\bibitem{korteweg1901forme}
D.~J. Korteweg, Sur la forme que prennent les {\'e}quations du mouvements des
  fluides si l'on tient compte des forces capillaires caus{\'e}es par des
  variations de densit{\'e} consid{\'e}rables mais connues et sur la
  th{\'e}orie de la capillarit{\'e} dans l'hypoth{\`e}se d'une variation
  continue de la densit{\'e}, Archives N{\'e}erlandaises des Sciences exactes
  et naturelles 6 (1901) 1--24.

\bibitem{montanero1997viscometric}
J.~M. Montanero, A.~Santos, Viscometric effects in a dense hard-sphere fluid,
  Physica A: Statistical Mechanics and its Applications 240~(1-2) (1997)
  229--238.

\bibitem{Gorji2014a}
M.~H. Gorji, {Fokker-Planck} solution algorithm for rarefied gas flows and
  applications of complex gas-surface interactions, Ph.D. thesis, Swiss Federal
  Institute of Technology in Zurich (ETHZ) (2014).

\bibitem{cercignani1988boltzmann}
C.~Cercignani, The {Boltzmann} equation, in: The {Boltzmann} equation and its
  applications, Springer, 1988, pp. 40--103.

\bibitem{frezzotti1997molecular}
A.~Frezzotti, Molecular dynamics and enskog theory calculation of one
  dimensional problems in the dynamics of dense gases, Physica A: Statistical
  Mechanics and its Applications 240~(1-2) (1997) 202--211.

\bibitem{kon2014method}
M.~Kon, K.~Kobayashi, M.~Watanabe, Method of determining kinetic boundary
  conditions in net evaporation/condensation, Physics of Fluids 26~(7) (2014)
  072003.

\bibitem{frezzotti2008comparison}
A.~Frezzotti, S.~Nedea, A.~Markvoort, P.~Spijker, L.~Gibelli, Comparison of
  molecular dynamics and kinetic modeling of gas-surface interaction, in: AIP
  Conference Proceedings, Vol. 1084, American Institute of Physics, 2008, pp.
  635--640.

\bibitem{Hirschfelder1963}
J.~Hirschfelder, R.~B. Bird, C.~F. Curtiss, Molecular theory of gases and
  liquids, Wiley, 1964.

\bibitem{carnahan1969equation}
N.~F. Carnahan, K.~E. Starling, Equation of state for nonattracting rigid
  spheres, The Journal of chemical physics 51~(2) (1969) 635--636.

\bibitem{oksendal2013stochastic}
B.~Oksendal, Stochastic differential equations: an introduction with
  applications, Springer Science \& Business Media, 2013.

\bibitem{Risken1989}
H.~Risken, The{ Fokker-Planck} Equation: Methods of solution and applications,
  Springer, 1989.

\bibitem{Gorji2011}
M.~H. Gorji, M.~Torrilhon, P.~Jenny, {Fokker--Planck} model for computational
  studies of monatomic rarefied gas flows, Journal of Fluid Mechanics 680
  (2011) 574--601.

\bibitem{Gorji2014}
M.~H. Gorji, P.~Jenny, An efficient particle {Fokker--Planck} algorithm for
  rarefied gas flows, Journal of Computational Physics 262 (2014) 325--343.

\bibitem{jenny2019accurate}
P.~Jenny, M.~H. Gorji, Accurate particle time integration for solving
  {Vlasov}-{Fokker}-{Planck} equations with specified electromagnetic fields,
  Journal of Computational Physics 387 (2019) 430--445.

\bibitem{pfeiffer2017adaptive}
M.~Pfeiffer, M.~H. Gorji, Adaptive particle--cell algorithm for
  {Fokker--Planck} based rarefied gas flow simulations, Computer Physics
  Communications 213 (2017) 1--8.

\bibitem{jun2019comparative}
E.~Jun, M.~Pfeiffer, L.~Mieussens, M.~H. Gorji, Comparative study between cubic
  and ellipsoidal {Fokker--Planck} kinetic models, AIAA Journal 57~(6) (2019)
  2524--2533.

\bibitem{li2013lattice}
Q.~Li, K.~Luo, X.~Li, Lattice boltzmann modeling of multiphase flows at large
  density ratio with an improved pseudopotential model, Physical Review E
  87~(5) (2013) 053301.

\bibitem{ghoufi2010calculation}
A.~Ghoufi, P.~Malfreyt, Calculation of the surface tension from multibody
  dissipative particle dynamics and monte carlo methods, Physical Review E
  82~(1) (2010) 016706.

\bibitem{elsner1991calculation}
A.~Elsner, Calculation of the surface tension according to van der waals,
  Physics Letters A 156~(3-4) (1991) 147--154.

\bibitem{elsner1989interaction}
A.~Elsner, Interaction potential of a saturated fluid, Physics Letters A
  138~(4-5) (1989) 168--172.

\bibitem{yasuoka1994evaporation}
K.~Yasuoka, M.~Matsumoto, Y.~Kataoka, Evaporation and condensation at a liquid
  surface. i. argon, The Journal of chemical physics 101~(9) (1994) 7904--7911.

\bibitem{kobayashi2016molecular}
K.~Kobayashi, K.~Hori, M.~Kon, K.~Sasaki, M.~Watanabe, Molecular dynamics study
  on evaporation and reflection of monatomic molecules to construct kinetic
  boundary condition in vapor--liquid equilibria, Heat and Mass Transfer 52~(9)
  (2016) 1851--1859.

\bibitem{pao1971application}
Y.-p. Pao, Application of kinetic theory to the problem of evaporation and
  condensation, The physics of Fluids 14~(2) (1971) 306--312.

\bibitem{frezzotti2003evidence}
A.~Frezzotti, P.~Grosfils, S.~Toxvaerd, Evidence of an inverted temperature
  gradient during evaporation/condensation of a lennard-jones fluid, Physics of
  fluids 15~(10) (2003) 2837--2842.

\bibitem{meland2003molecular}
R.~Meland, Molecular dynamics simulation of the inverted temperature gradient
  phenomenon, Physics of Fluids 15~(10) (2003) 3244--3247.

\bibitem{meland2004dependence}
R.~Meland, T.~Ytrehus, Dependence of the inverted temperature gradient
  phenomenon on the condensation coefficient, Physics of Fluids 16~(3) (2004)
  836--838.

\bibitem{gorji2013fokker}
M.~H. Gorji, P.~Jenny, A {Fokker--Planck} based kinetic model for diatomic
  rarefied gas flows, Physics of fluids 25~(6) (2013) 062002.

\bibitem{shan1993lattice}
X.~Shan, H.~Chen, Lattice {Boltzmann} model for simulating flows with multiple
  phases and components, Physical review E 47~(3) (1993) 1815.

\bibitem{watanabe2012phase}
H.~Watanabe, N.~Ito, C.-K. Hu, Phase diagram and universality of the
  lennard-jones gas-liquid system, The Journal of chemical physics 136~(20)
  (2012) 204102.

\bibitem{siggia1979late}
E.~D. Siggia, Late stages of spinodal decomposition in binary mixtures,
  Physical review A 20~(2) (1979) 595.

\bibitem{bastea1997spinodal}
S.~Bastea, J.~Lebowitz, Spinodal decomposition in binary gases, Physical review
  letters 78~(18) (1997) 3499.

\bibitem{fasoulas2019}
S.~Fasoulas, C.-D. Munz, M.~Pfeiffer, J.~Beyer, T.~Binder, S.~Copplestone,
  A.~Mirza, P.~Nizenkov, P.~Ortwein, W.~Reschke, Combining particle-in-cell and
  direct simulation monte carlo for the simulation of reactive plasma flows,
  Physics of Fluids 31~(7) (2019) 072006.

\bibitem{pfeiffer2019particle}
M.~Pfeiffer, F.~Hindenlang, T.~Binder, S.~Copplestone, C.-D. Munz, S.~Fasoulas,
  A particle-in-cell solver based on a high-order hybridizable discontinuous
  galerkin spectral element method on unstructured curved meshes, Computer
  Methods in Applied Mechanics and Engineering 349 (2019) 149--166.

\bibitem{cockburn2009hybridizable}
B.~Cockburn, B.~Dong, J.~Guzm{\'a}n, M.~Restelli, R.~Sacco, A hybridizable
  discontinuous galerkin method for steady-state convection-diffusion-reaction
  problems, SIAM Journal on Scientific Computing 31~(5) (2009) 3827--3846.

\bibitem{arnold2002unified}
D.~N. Arnold, F.~Brezzi, B.~Cockburn, L.~D. Marini, Unified analysis of
  discontinuous galerkin methods for elliptic problems, SIAM journal on
  numerical analysis 39~(5) (2002) 1749--1779.

\end{thebibliography}

\section*{Funding}
\noindent 
This work was supported by the Deutscher Akademischer Austauschdienst (M.S., grant numbers 57438025); the Swiss National Science Foundation (H.G., grant number 174060); and the Deutsche Forschungsgemeinschaft (M.P., grant number 393159129).


\section*{Declaration of Interests}
The authors report no conflict of interest.

\section*{Supplementary data}
 Supplementary material and movies are available at https://doi.org/10.1017/jfm.2019.

\end{document}